\documentclass[onecolumn,12pt]{article}
\usepackage{jheppub}
\usepackage{ifpdf}
\usepackage[bb=boondox]{mathalfa}
\usepackage{graphicx,subcaption}
\graphicspath{ {figures/} }
\usepackage{amsfonts}
\usepackage{amsmath,braket}
\usepackage{amssymb}
\usepackage{epsfig}
\usepackage{tensor}
\usepackage{pdfpages}
\usepackage{bbm}
\usepackage{graphicx,epstopdf}
\usepackage[numbers]{natbib} 
\usepackage[makeroom]{cancel}
\usepackage{hyperref}
\usepackage{array}
\usepackage[export]{adjustbox}

\usepackage[normalem]{ulem}
\usepackage{romannum}

\usepackage{ulem}

\usepackage{color}
\usepackage{here}

\numberwithin{equation}{section}									% equation numbering by section

\usepackage{comment}

\newcommand{\de}{\partial}
\newcommand{\be}{\begin{equation}}
	\newcommand{\ba}{\begin{eqnarray}}
		\newcommand{\ea}{\end{eqnarray}}
	\newcommand{\ee}{\end{equation}}

\newcommand{\s}{\sqrt}

\newcommand{\ti}{\tilde}

\newcommand{\ddd}{\cdot\cdot\cdot}
\newcommand{\no}{\nonumber \\}

\newcommand{\la}{\langle}
\newcommand{\lb}{\rangle}
\newcommand{\bea}{\begin{eqnarray}}
	\newcommand{\eea}{\end{eqnarray}}
\newcommand{\bes}{\begin{equation*}}
	\newcommand{\beas}{\begin{eqnarray*}}
		\newcommand{\eeas}{\end{eqnarray*}}
	\newcommand{\bas}{\begin{array*}}
		\newcommand{\eas}{\end{array*}}
	\newcommand{\ees}{\end{equation*}}
\newcommand{\nn}{\nonumber}

\newcommand{\ep}{\epsilon}

         \let\l=\lambda  

\def\nn{\nonumber}

\newcommand{\arctanh}{\text{arctanh}}

\subheader{\today}
\title{\boldmath Entropic Interpretation of Einstein Equation in dS/CFT}

\author[a]{Kosei Fujiki,}
\author[a]{Michitaka Kohara,}
\author[a]{Kotaro Shinmyo,}
\author[a]{Yu-ki Suzuki,}
\author[a,b]{Tadashi Takayanagi}

\affiliation[a]{Center for Gravitational Physics and Quantum Information (CGPQI), \\
Yukawa Institute for Theoretical Physics (YITP), Kyoto University,\\
	Kitashirakawa Oiwakecho, 
    Sakyo-ku, Kyoto 606-8502, Japan}
\affiliation[b]{Inamori Research Institute for Science,\\
	620 Suiginya-cho, Shimogyo-ku,Kyoto 600-8411 Japan}

\emailAdd{kosei.fujiki@yukawa.kyoto-u.ac.jp}
\emailAdd{michitaka.kohara@yukawa.kyoto-u.ac.jp}
\emailAdd{kotaro.shinmyo@yukawa.kyoto-u.ac.jp}
\emailAdd{yu-ki.suzuki@yukawa.kyoto-u.ac.jp}
\emailAdd{takayana@yukawa.kyoto-u.ac.jp}

\abstract{In this paper, we demonstrate that the first law of holographic pseudo-entropy, which is a non-Hermitian generalization of entanglement entropy in a two-dimensional conformal field theory (CFT), is equivalent to the perturbative Einstein equation in three-dimensional de Sitter (dS) space, assuming the dS/CFT correspondence. Our analysis reveals that the geodesic that accurately satisfies the first law of holographic pseudo-entropy consists of a timelike curve and a curve whose coordinates are complex. We also demonstrate that infinitesimal changes to the pseudo entropy satisfy a Klein–Gordon equation in two-dimensional de Sitter space. These imply the emergence of a time coordinate from a Euclidean CFT in dS/CFT. }

\begin{document} 
	
	%%%%%%%%%%%%%%%%%%%%%%%%%%%%%%%%%%%%%%%%%%%%%%%%%%%%
	\begin{flushright}
		YITP-25-173
        \\
	\end{flushright}
	%%%%%%%%%%%%%%%%%%%%%%%%%%%%%%%%%%%%%%%%%%%%%%%%%%%%
	\maketitle
	\flushbottom

%%%%%%%%%%%%%%%%%%%%%%%%%%%%%%%%%%%%%%%%%%%%%%%%%%%%%%%%
%%%%%%%%%%%%%%%%%%%%%%%%%%%%%%%%%%%%%%%%%%%%%%%%%%%%%%%%
%\section{Introduction}
%\label{sec:intro}
%%%%%%%%%%%%%%%%%%%%%%%%%%%%%%%%%%%%%%%%%%%%%%%%%%%%%%%%
%%%%%%%%%%%%%%%%%%%%%%%%%%%%%%%%%%%%%%%%%%%%%%%%%%%%%%%%

%\begin{figure}[htbp]
%		\centering
%		\includegraphics[width=13cm]{setup.pdf}
%		\caption{} 
%		\label{fig:setup}
%\end{figure}

%%%%%%%%%%%%%%%%%%%%%%%%%%%%%%%%%%%%%%%%%%%%%%%%%%%%%%%%
%%%%%%%%%%%%%%%%%%%%%%%%%%%%%%%%%%%%%%%%%%%%%%%%%%%%%%%%
\section{Introduction}
\label{sec:Intro}
%%%%%%%%%%%%%%%%%%%%%%%%%%%%%%%%%%%%%%%%%%%%%%%%%%%%%%%%
%%%%%%%%%%%%%%%%%%%%%%%%%%%%%%%%%%%%%%%%%%%%%%%%%%%%%%%%

One of the ultimate goals of theoretical physics will be to construct a satisfactory theory of quantum gravity which describes the creation of the Universe. Even though we can well describe a macroscopic spacetime by using general relativity, we cannot handle a Planck scale one due to strong quantum gravity effects. The classical geometry description of a gravitational spacetime is no longer valid for the quantum Universe. For example, we do not even know how to define the size of quantum Universe within the conventional framework of gravity. To make further progress, we need to explore new approaches to develop our understanding of quantum gravity. 

One of the most promising approaches to quantum gravity is the 
holographic duality, which allows us to relate gravitational theories to various theories of quantum many-body systems which live on their boundaries \cite{tHooft:1993dmi,Susskind:1994vu}. The most successful example is the AdS/CFT correspondence \cite{Maldacena:1997re,Gubser:1998bc,Witten:1998qj} and this argues that a quantum gravity on a ($d+1$)-dimensional AdS space (AdS$_{d+1}$) is equivalent to a $d$-dimensional conformal field theory (CFT$_d$). In AdS/CFT, it has been expected that the geometry of gravitational spacetime may emerge from quantum entanglement \cite{Swingle:2009bg,VanRaamsdonk:2010pw}, motivated by the holographic calculation of entanglement entropy in terms of areas of extremal surfaces \cite{Ryu:2006bv,Ryu:2006ef,Hubeny:2007xt}. One important evidence of this idea is the derivation of the perturbative Einstein equation in the AdS$_{d+1}$ from the first law of entanglement entropy in CFT$_d$ \cite{Lashkari:2013koa,Faulkner:2013ica}.
The perturbative Einstein equation turns out to be equivalent to a special differential equation \cite{Nozaki:2013vta,Bhattacharya:2013bna,deBoer:2015kda,deBoer:2016pqk} satisfied by a small perturbation of  holographic entanglement entropy, which reflects the structure called kinematical spaces \cite{Czech:2015qta,Czech:2016xec}.
This differential equation can also be derived from the first law of entanglement entropy \cite{Blanco:2013joa,Wong:2013gua,Bhattacharya:2012mi}, which is universal for a CFT$_d$. Moreover, the equivalence between the Einstein equation on an AdS$_{d+1}$ and the entanglement entropy in a CFT$_d$ have been shown for higher orders of the perturbations \cite{Faulkner:2017tkh,Sarosi:2017rsq}.

In a gravity which describes a macroscopic spacetime, its spatial size can be estimated by the holographic entanglement entropy as it is proportional to areas of extremal surfaces. Even when the size of spacetime is microscopic, we may employ the entanglement entropy to define its size. This motivates us to study the quantum cosmology in terms of its dual CFT by focusing on the dynamics of entanglement entropy. In quantum field theories with a suitable UV regularization or more generally quantum many-body systems, the entanglement entropy is one of the most fundamental and robust quantities which we can rely on. The main purpose of this paper is to develop these ideas in the context of holographic duality for a de Sitter space (dS), so-called dS/CFT correspondence \cite{Strominger:2001pn,Maldacena:2002vr,Witten:2001kn}. dS is the most basic cosmological spacetime, having in mind the problem of understanding the mechanism of creating the Universe. 

In the dS/CFT, a ($d+1$)-dimensional dS (dS$_{d+1}$) is argued to be equivalent to a $d$-dimensional Euclidean CFT (CFT$_{d}$) on its boundary. In this holography, the real time direction should emerge from the CFT without any time, which makes the dual CFT very exotic. Indeed the dual CFT is known to be non-unitary, which can be immediately seen from the fact that the central charges when $d=$even take pure imaginary values in the classical gravity limit \cite{Maldacena:2002vr}. This fact can also be understood from the explicit examples of dS$_{d+1}/$CFT$_{d}$. For $d=3$, a CFT$_3$ dual to higher spin gravity for dS$_4$ was given in terms of SP$(N)$ model \cite{Anninos:2011ui}, which is a field theory with ghost fields. For $d=2$, a CFT$_2$ dual to the Einstein gravity on dS$_3$ was found in terms of a non-rational extension of the SU$(2)$ WZW \cite{Hikida:2021ese}, which is equivalent to the Liouville theory with a complex value of the central charge \cite{Hikida:2022ltr}. Interestingly, the latter CFT also appears in the context of the proposed holographic duality between the two-sided double scaled SYK model and the static patch of dS$_3$ recently proposed in  \cite{Verlinde:2024zrh,Blommaert:2025eps}. In the dS$_3/$CFT$_2$, the holographic entanglement entropy takes complex values and this should be interpreted as the pseudo entropy \cite{Nakata:2021ubr} more properly (for other proposals, see \cite{Ruan:2025uhl,Sanches:2016sxy,Arias:2019pzy}). This is because the density matrix for this non-unitary CFT turns out to be non-Hermitian \cite{Doi:2022iyj,Doi:2023zaf}.  
In this paper, we would like to study how the Einstein equation is encoded in the dynamics of the holographic pseudo entropy, which is a dS/CFT counterpart of holographic entanglement entropy. We focus on the dS$_3/$CFT$_2$ and show that the first law of pseudo entropy \cite{Mollabashi:2021xsd} for the non-unitary Euclidean CFT is equivalent to the perturbative Einstein equation for the dS$_3$ gravity. In this analysis we pay attention to the role played by the Lorentzian time in dS$_3$, which is expected to emerge from the Euclidean CFT. We will show that we can obtain desirable results by extending geodesics whose lengths compute the holographic pseudo entropy
to those in complexified spacetimes. Similar complexifications of extremal surfaces or geodesics were already implemented and studied in \cite{Heller:2024whi,Heller:2025kvp,Nunez:2025ppd} in the context of time-like entanglement entropy \cite{Doi:2022iyj,Doi:2023zaf} (see also 
\cite{Narayan:2022afv,Foligno:2023dih,Narayan:2023ebn,Chu:2023zah,Grieninger:2023knz,Guo:2024lrr,Milekhin:2025ycm,Chu:2025sjv,Bou-Comas:2024pxf,Xu:2024yvf,Nunez:2025gxq,Nunez:2025puk}). Moreover, we show that the infinitesimally small change of pseudo entropy satisfies a partial differential equation which corresponds to a scalar field on dS$_2$, which implies the emergent time from an Euclidean CFT$_2$.

The contents of this paper are as follows. In section two, we first present a brief review of the first law of entanglement entropy in CFTs and the equivalence between the perturbative Einstein equation and the first law of entanglement entropy for the Poincar\'{e} AdS setup. 
In section three, we analyze the holographic pseudo entropy in dS$_3$ under a small perturbation of the pure dS$_3$, under the real space ansatz of geodesic. We show that this does not reproduce the expected first law of pseudo entropy. In section four, we employ an analytical continuation from AdS$_3$ to dS$_3$ to compute the infinitesimal change of the holographic pseudo entropy.
This leads to geodesics in complex valued coordinates.
We show that the perturbative Einstein equation is equivalent to the first law of holographic pseudo entropy. We present how the geodesic looks like in our dS/CFT example.
In section five, we summarize our conclusions and discuss future problems. In appendix A, we give a derivation of a Banados-like map for dS$_3/$CFT$_2$, which leads to another candidate of geodesic.

%%%%%%%%%%%%%%%%%%%%%%%%%%%%%%%%%%%%%%%%%%%%%%%%%%%%%%%%
%%%%%%%%%%%%%%%%%%%%%%%%%%%%%%%%%%%%%%%%%%%%%%%%%%%%%%%%
\section{First law and AdS/CFT}
\label{sec:1st}
%%%%%%%%%%%%%%%%%%%%%%%%%%%%%%%%%%%%%%%%%%%%%%%%%%%%%%%%
%%%%%%%%%%%%%%%%%%%%%%%%%%%%%%%%%%%%%%%%%%%%%%%%%%%%%%%%
As a preparation of our analysis of the dS/CFT in the next section, here we first give a brief review of the equivalence between the  perturbative Einstein equation in AdS and the first law of entanglement entropy in its dual CFT. Then we will generalize the result to  the hyperbolic ball coordinate of Euclidean AdS.

\subsection{First law of entanglement entropy} 
In thermodynamics, the variation of thermal entropy $S$ is related to that of the energy $E$ via the first law of thermodynamics under a fixed temperature $T$, given by the well-known relation $\Delta E=T\cdot\Delta S$.
A similar relation is also known to hold for the entanglement entropy. 

Consider a quantum state $|\Psi\lb$ which describes the total system, whose Hilbert space $H$ is factorized as $H=H_A\otimes H_B$, where $H_A$ and $H_B$ are Hilbert spaces for subsystems $A$ and $B$. The entanglement entropy $S_A$ is defined as the von-Neumann entropy $S_A=-\mbox{Tr}_A[\rho_A\log\rho_A]$ for the reduced density matrix $\rho_A=\mbox{Tr}_B[|\Psi\lb\la\Psi|]$, which describes the state of an observer who is only accessible to $A$. We introduce the modular Hamiltonian by $H_A=-\log \rho_A$ and write its expectation value as $E_A=\la H_A\lb$. In this setup, we can show the following 
relation, so-called first law of entanglement entropy \cite{Blanco:2013joa,Wong:2013gua} (see also \cite{Bhattacharya:2012mi} in the case of small subsystems):
\ba
\Delta S_A\simeq \Delta E_A, \label{firstLA}
\ea
for an infinitesimally small variation of the state $|\Psi\lb$.

Consider infinitesimally small excitations around the vacuum in a $d$-dimensional CFT. We take its coordinate to be $(x_0,x_1,x_2,\ddd,x_{d-1})$.
We choose the subsystem $A$ to be a ($d-1$)-dimensional round ball $|x-\xi|\leq l$ on the time slice $x_0=\xi_0$, where $\xi_i\ \ (i=1,2,\ddd,d)$ is the center of the ball and $l$ is its radius. In this setup, as derived in \cite{Blanco:2013joa}, we can show that the modular Hamiltonian becomes local and given explicitly as 
\be
 H_A=2\pi \int_{|x-\xi|\leq l}d^{d-1}x\frac{l^2-|x-\xi|^2}{2l}
T_{tt}(x),\label{FEE}
\ee
where $T_{tt}$ is the energy density. 
Via the Fourier transformation $\xi\to k$, this is rewritten as \cite{Blanco:2013joa}
\ba
\Delta S_A\propto \frac{l^{\frac{d-1}{2}}}{|k|^{\frac{d+1}{2}}}J_{\frac{d+1}{2}}(l|k|)\cdot T_{tt}(k,l). \label{firstpert}
\ea

\subsection{Einstein equation in AdS and first law of entanglement entropy}

Now let us briefly review the equivalence between the perturbative Einstein equation in the AdS$_{d+1}$ and the first law of entanglement entropy for a CFT$_{d}$. The holographic entanglement entropy \cite{Ryu:2006bv,Ryu:2006ef,Hubeny:2007xt} argues that the entanglement entropy in the dual CFT$_d$ can be computed from the area of ($d-1$)-dimensional extremal surface $\Gamma_A$ in the bulk AdS$_{d+1}$, which ends on the boundary of subregion $A$ by the formula
\ba
S_A=\frac{\mbox{Area}(\Gamma_A)}{4G_N}.  \label{RT}
\ea
We consider an infinitesimal perturbation $h_{\mu\nu}$ of the metric around the Poincare AdS$_{d+1}$, which looks like
\ba
ds^2=G_{ab}dy^ady^b=\frac{dz^2+(\eta_{\mu\nu}+h_{\mu\nu})dx^\mu dx^\nu}{z^2},  \label{pertr}
\ea
where $y=(z,x)$ is the coordinate of the bulk AdS$_{d+1}$. We set the AdS radius to be unit throughout this paper.
We focus on the most symmetric choice of subregion $A$ for the CFT$_d$, namely the $d-1$ dimensional round ball on a time slice $x_0=\xi_0$, with the radius $l$ and its center at $(x_1,x_2,\ddd,x_{d-1})=(\xi_1,\xi_2,\ddd,\xi_{d-1})$.

Before the perturbation $h_{\mu\nu}=0$, the extremal surface which computes entanglement entropy $S_A$ via (\ref{RT}) is given by the round semi-sphere given by 
\ba
z^2+\sum_{i=1}^{d-1}(x_i-\xi_i)^2=l^2.   \label{minsp}
\ea
at $x_0=\xi_0$. The first order perturbation to the area of the extremal surface $\Gamma_A$ is insensitive to the deformation of the extremal surface itself because the first order variation of the area functional vanished due to the extremal surface condition. Therefore, we can evaluate the first order perturbation to $S_A$ by using the same surface (\ref{minsp}) and by taking into account the metric perturbation as follows:
\ba
\Delta S_A\simeq \frac{1}{8G_N}\int_{\Gamma_A} d^{d-1}y \s{G^{(0)}}G^{(1)}_{ab} G^{(0)ab},\label{CHEE}
\ea
where $G^{(0)}$ is the pure AdS$_{d+1}$ metric and $G^{(1)}$ is its first order perturbation, which can be found from 
(\ref{pertr}) straightforwardly. 

We require that  the metric $G_{ab}$ satisfies the vacuum Einstein equation with the negative cosmological constant:
\ba
R_{ab}-\frac{1}{2}R\,G_{ab}-\frac{d(d-1)}{2}G_{ab}=0.
\ea
After reasonable computations, we can rewrite this Einstein equation for the infinitesimally small perturbation $G^{(1)}$ (\ref{pertr}) into the following differential equation, satisfied by the extremal surface or, equally, the holographic entanglement entropy  \cite{Nozaki:2013vta,Bhattacharya:2013bna}:
\ba
\left[\frac{\de^2}{\de l^2}-\frac{(d-2)}{l}\frac{\de}{\de l}-\frac{d}{l^2}-\sum_{i=1}^{d-1}\frac{\de^2}{\de \xi_i^2}\right]\Delta S_A=0. \label{EOM}
\ea
Indeed, it is a simple exercise to confirm that (\ref{firstpert}) satisfies (\ref{EOM}). Moreover, the consequence of Einstein equation (\ref{EOM}) and the well-known relation between the holographic energy stress tensor 
and the behavior of the metric near the AdS boundary precisely reproduce the first law (\ref{EOM}) or equally (\ref{FEE}), where the latter leads to 
$\Delta S_A\propto l^d T_{tt}(k,l)$ in the small subsystem limit $l\to 0$ \cite{Bhattacharya:2012mi}.
In this way, the perturbative Einstein equation is equivalent to the first law of EE (\ref{FEE}) \cite{Lashkari:2013koa,Faulkner:2013ica}. This relation is extended to higher order with matter fields in \cite{Faulkner:2017tkh,Sarosi:2017rsq}. See \cite{VanRaamsdonk:2016exw} for a review.

It is also useful to note that the differential equation (\ref{EOM}) can be expressed as the massive scalar field equation on a $d$-dimensional de-Sitter space $dS_d$ \cite{deBoer:2015kda}:
\ba
(\Box_{dS_d}+d)\Delta S_A=0,  \label{FSEOM}
\ea
where we treat $\Delta S_A$ as a scalar field and 
$\Box_{dS_d}$ is the Laplacian on a dS$_d$ whose metric reads 
\ba
ds^2=\frac{-dl^2+\sum_{i=1}^{d-1}d\xi^2_i}{l^2}. \label{dSme}
\ea

\subsection{CFT$_2$ Case}

Let us explicitly examine the relation in the $d=2$ case, namely AdS$_3/$CFT$_2$. In this case, the full analytical expression of entanglement entropy is available. It is well-known that the entanglement entropy for an interval $A$ whose end points are $(x^+,x^-)$ and $(y^+,y^-)$ reads
\ba
S_A=S_R+S_L,
\label{2dCFTEE}
\ea
where
\ba
&& S_R=\frac{c}{12}\log \frac{(f(x^+)-f(y^+))^2}{f'(x^+)f'(y^+)\ep^2},
\ \ \ \ \  S_L=\frac{c}{12}\log \frac{(g(x^-)-g(y^-))^2}{g'(x^-)g'(y^-)\ep^2},
\ea
as obtained from the conformal mapping $x^{+'}=f(x^+)$ and $x^{-'}=g(x^-)$.
As found in \cite{deBoer:2016pqk}, this satisfies the Liouville equations
\ba
&& \frac{\de}{\de x^+}\frac{\de}{\de y^+}S_R=\frac{c}{6\ep^2}e^{-\frac{12}{c}S_R},\ \ \ \ \  \frac{\de}{\de x^-}\frac{\de}{\de y^-}S_L=\frac{c}{6\ep^2}e^{-\frac{12}{c}S_L}, \label{ADStheq}
\ea
where $\ep$ is the UV cut off.

Let us focus on a time slice $t=t_0$ such that $x^{\pm}=x\pm t_0$
and  $y^{\pm}=y\pm t_0$. By summing the two equations in (\ref{ADStheq}) we obtain
\ba
&& \frac{\de}{\de x}\frac{\de}{\de y}S_A-\frac{c}{6\ep^2}\left(e^{-\frac{12}{c}S_R}+e^{-\frac{12}{c}S_L}\right)=0.
\label{ADStoto}
\ea
Now we assume that the functions $f$ and $g$ are very close to the identity, 
i.e., 
\be
f(x)=x+\delta f(x),\ \ \ \ \ g(x)=x+\delta g(x), \label{infperty}
\ee
where we treat $\delta f$ and $\delta g$ as infinitesimal small perturbations. At the first order of this perturbation, we can rewrite (\ref{ADStoto}) as follows
\ba
\left(\frac{\de^2}{\de x\de y}+\frac{2}{(y-x)^2}\right)\Delta S_A=0,\label{FDEQ}
\ea
where the perturbation of entanglement entropy is defined by $\Delta S_A=S_A-\frac{c}{6}\log\frac{(y-x)^2}{\ep^2}$. As in (\ref{dSme}) we introduce the dS$_2$ metric as $ds^2=\frac{4dxdy}{(x-y)^2}$.
Then (\ref{FDEQ}) can be expressed as $(\Box_{dS}+2)\Delta S_A=0$ and this coincides with (\ref{FSEOM}) at $d=2$.

It is also straightforward to confirm the first law (\ref{FEE}) directly.
By the infinitesimal perturbation (\ref{infperty}) we find
\ba
&&\Delta S_R\simeq \frac{c}{12}\left[-\delta f'(x)-\delta f'(y)\right]+\frac{c}{6}\cdot\frac{\delta f(y)-\delta f(x)}{y-x},\no
&&\Delta S_L\simeq \frac{c}{12}\left[-\delta g'(x)-\delta g'(y)\right]+\frac{c}{6}\cdot\frac{\delta g(y)-\delta g(x)}{y-x}.
\ea
The energy density is given by the Schwarzian derivative as usual and thus reads 
\ba
&& T_{tt}=\frac{c}{12\pi}\left[\frac{3(f'')^2-2f'f'''}{4(f')^2}+\frac{3(g'')^2-2g'g'''}{4(g')^2}\right]\simeq 
-\frac{c}{24\pi}\delta f'''-\frac{c}{24\pi}\delta g'''.\label{SH}
\ea
By repeating partial integrations twice, we can show 
\ba
&& \int^y_x d\xi\frac{\left(\frac{y-x}{2}\right)^2-\left(\xi-\frac{y+x}{2}\right)^2}{y-x}  T_{tt}(\xi)\no
&&=-\frac{c}{12\pi(y-x)}\int^y_{x}d\xi\left(\xi-\frac{x+y}{2}\right)\left(\delta f''(\xi)+\delta g''(\xi)\right)\no
&&= -\frac{c}{24\pi}\left[\delta f'(y)+\delta g'(x)+\delta g'(y)+\delta f'(x)-2\frac{\delta f(y)-\delta f(x)+\delta g(y)-\delta g(x)}{y-x}\right].\no
\ea
Thus this reproduces the first law relation (\ref{firstLA}) for $d=2$.

%%%%%%%%%%%%%%%%%%%%%%%%%%%%%%%%%%%%%%%%%%%%%%%%%%%%%%%%
%%%%%%%%%%%%%%%%%%%%%%%%%%%%%%%%%%%%%%%%%%%%%%%%%%%%%%%%
\section{First law and Perturbations in dS$_3$}
\label{sec:dsCFT}
%%%%%%%%%%%%%%%%%%%%%%%%%%%%%%%%%%%%%%%%%%%%%%%%%%%%%%%%
%%%%%%%%%%%%%%%%%%%%%%%%%%%%%%%%%%%%%%%%%%%%%%%%%%%%%%%%
Now, we move on to the main problem of this paper, namely the relation between the first law of entanglement entropy and the dS/CFT. We focus on the $d=2$ case, i.e., dS$_3/$CFT$_2$ for simplicity. It is known that the holographic entanglement entropy generally takes complex values in the dS/CFT \cite{Narayan:2015vda,Sato:2015tta,Hikida:2022ltr,Doi:2022iyj,Narayan:2022afv,Doi:2023zaf,Narayan:2023ebn}. This complex valued entropy is properly interpreted as the pseudo entropy \cite{Nakata:2021ubr}, which is a generalization of entanglement entropy for the cases where the density matrices are not Hermitian. Moreover, its geometric calculation in terms of bulk dS was proposed in  \cite{Doi:2022iyj,Doi:2023zaf} by extending the holographic entanglement entropy formula (\ref{RT}) to the cases when a part of the extremal surface becomes time-like, which produces the imaginary part of the pseudo entropy.
Below, we examine how the holographic pseudo entropy is changed by infinitesimal perturbations of dS$_3$. Though under this deformation, we expect that the first law relation is still correct as in the AdS case, this looks difficult at first sight as we will see in this section, because the entropy takes complex values in the dS/CFT case. However, as we will show in the next section, this works well if we admit the new feature that a geodesic in a complexified dS$_3$ contributes.

\subsection{First law in the dual CFT$_2$}
The dS$_3/$CFT$_2$ argues that a gravitational theory on a dS$_3$ is dual to a CFT$_2$, which lives on the boundary of dS$_3$, i.e., the future infinity \cite{Strominger:2001pn,Maldacena:2002vr}. If we consider the Einstein gravity on a dS$_3$, the dual two dimensional CFT is expected to be non-unitary as its central charge $c$ is given by the imaginary value
\ba
c=i\frac{3R_{dS}}{2G_N},
\ea
where $R_{dS}$  is the dS radius, which is set to $R_{dS}=1$ below. It is also known that the energy density $T_{tt}$ can take 
imaginary values. Nevertheless, we expect that the first law of entanglement entropy remains correct though it should be called the first law of pseudo entropy as first considered in \cite{Mollabashi:2021xsd}. When we choose the subsystem $A$ to be the interval $x_1\leq x \leq x_2$ at a fixed time $t=t_0$, the first law of pseudo entropy reads
\ba
    \Delta S_A = 2\pi \int_{x_1}^{x_2} dx \frac{(\frac{x_2-x_1}{2})^2-(x-\frac{x_2+x_1}{2})^2}{x_2-x_1} T_{tt}(t_0,x).
    \label{eq:modHonplane}
\ea
Indeed, this expression is independent of the central charge and 
it is natural that it takes the same form even for the dS/CFT.
Note that in the CFT$_2$ dual to dS$_3$, though $\Delta S_A$ becomes imaginary, this is consistent if $T_{tt}$ also becomes imaginary. We will also give a formal derivation of (\ref{eq:modHonplane}) from CFT calculations in the next subsection (\ref{eq:modHonplane}).

A standard setup of dS/CFT is for the global coordinates of a de Sitter space. The future boundary of this space is given by a sphere. Thus, it is useful to present the first law of pseudo entropy for a CFT$_2$ on S$^2$. We take the coordinate $(\phi, \theta)$ such that the metric of S$^2$ reads $ds^2=d\theta^2+\sin^2\theta d\phi^2$. We chose the subregion $A$ as the interval $\theta_1\leq\theta\leq\theta_2$ at a fixed value of Euclidean time $\phi=\phi_*$. Then, by analyzing  the conformal transformation $x+it_E=e^{i\phi}\cot\frac{\theta}{2}$
from the planar coordinates to the spherical coordinates using the stereographic projection, we finally obtain the following formula for the first law:
\ba
    \Delta S_A 
%=&& 2\pi \int^{\theta_2}_{\theta_1}  d\theta   
%    \frac{-\tan^2\frac{\theta}{2}}{1-%\cos\theta}\frac{\left(\frac{\frac{\sin \theta_2 }{1-\cos %\theta_2 }-\frac{\sin \theta_1
%   }{1-\cos \theta_1 }}{2}\right)^2-\left(\frac{\sin \theta}{1-%\cos \theta}-\frac{\left(\frac{\sin \theta_1 }{1-\cos \theta_1 %}+\frac{\sin
%   \theta_2 }{1-\cos \theta_2 }\right)}{2}\right)^2}{\frac{\sin %\theta_2 }{1-\cos \theta_2 }-\frac{\sin \theta_1 }{1-\cos %\theta_1 }}
%    T_{\phi\phi} \no
    =&& 2\pi\int^{\theta_2}_{\theta_1}d\theta \frac{\tan^2\frac{\theta}{2}}{2\sin^2\left(\frac{\theta}{2}\right)}\cdot \frac{\sin\left(\frac{\theta-\theta_1}{2}\right)\sin\left(\frac{\theta_2-\theta}{2}\right)}{\sin^2\left(\frac{\theta}{2}\right)\sin\left(\frac{\theta_2-\theta_1}{2}\right)}   T_{\phi\phi}.
    \label{modHonsphere}
\ea
Below, we would like to discuss how we can reproduce this CFT expectation from the dual gravity calculations. 

\subsection{CFT$_2$ derivation of first law}\label{sec:cftt}
Now, we present an explicit calculation of the entanglement entropy in the CFT$_2$, which is expected to be dual to dS$_3$. The CFT$_2$ is defined on a complex plane whose coordinate is written as $(w,\bar{w})$ and has the imaginary valued central charge $c=i\ti{c}$, where $\ti{c}$ is a positive constant. We choose the subregion $A$ whose end points are $(w_1,\bar{w_1})$ and $(w_2,\bar{w_2})$. 
% The corresponding subregion A' is given between $(\xi_1,\bar{\xi_1})$ and $(\xi_2,\bar{\xi_2})$ in the Poincar\'{e} dS$_3$.
Considering the conformal transformation \eqref{eq:conftrsf}, we obtain the following entanglement entropy:
\begin{equation}
    S_A = S+\bar{S}+\frac{\pi \tilde{c}}{6} ,
    \label{eq:SAdS}
\end{equation}
where
\ba
    S=\frac{i\tilde{c}}{12}\log\frac{\left(f(w_1)-f(w_2)\right)^2}{f'(w_1)f'(w_2)\ep^2},\;\;\; \bar{S}=\frac{i\tilde{c}}{12}\log\frac{\left(\bar{f}(\bar{w}_1)-\bar{f}(\bar{w}_2)\right)^2}{\bar{f}'(\bar{w}_1)\bar{f}'(\bar{w}_2)\ep^2}.
\ea
Here we noted that the central charge of the CFT is pure imaginary $c=i\ti{c}$ and the cut off $\ep$  differs from that of the ordinary CFT$_2$ dual to AdS$_3$ by a factor of $i$ as argued in \cite{Doi:2022iyj,Doi:2023zaf}, which leads to the constant term $\frac{\pi\ti{c}}{6}$.

As is the previous case, we can find the Liouville equations
\ba
    \frac{\de}{\de w_1}\frac{\de}{\de w_2}S=\frac{i\tilde{c}}{6\ep^2}e^{-\frac{12}{i\tilde{c}}S}, \;\;\; \frac{\de}{\de \bar{w}_1}\frac{\de}{\de \bar{w}_2}\bar{S}=\frac{i\tilde{c}}{6\ep^2}e^{-\frac{12}{i\tilde{c}}\bar{S}}.
    \label{eq:LiouvilledS}
\ea
We focus on the Euclidean time slice $y=y_0$.  For the infinitesimally small conformal transformation
\ba
    \xi=w+\delta f(w), \qquad \bar{\xi}=\bar{w}+\delta \bar{f}(\bar{w}),
\ea
we obtain
\ba
    \left(\frac{\de}{\de x_1}\frac{\de}{\de x_2}+\frac{2}{(x_1-x_2)^2}\right)\Delta S_A=0,
    \label{perturbedLiouvilledS}
\ea
where $\Delta S_A=S_A-\left(\frac{i\tilde{c}}{6}\log\frac{(x_1-x_2)^2}{\ep^2}+\frac{\pi\tilde{c}}{6}\right)$. This is identical to (\ref{FDEQ}) and thus we have the same differential equation $(\Box_{dS}+2)\Delta S_A=0$, introducing the dS$_2$ metric as $ds^2=\frac{4dx_1dx_2}{(x_1-x_2)^2}$. 

The first order perturbed entanglement entropy is evaluated as
\ba
    \Delta S_A \simeq \frac{i\tilde{c}}{6}\Bigg(\frac{\delta f(w_2)+\delta \bar{f}(\bar{w}_2)-\delta f(w_1)-\delta \bar{f}(\bar{w}_1)}{x_2-x_1}
     -\frac{\delta f'(w_2)+\delta \bar{f}'(\bar{w}_2)+\delta f'(w_1)+\delta \bar{f}'(\bar{w}_1)}{2}\Bigg). \notag \\
    \label{eq:dSDeltaS}
\ea
The energy stress tensor can also be obtained from the
schwarzian derivative as in (\ref{SH}), which leads to 
\ba
    T_{yy} \simeq-\frac{i \tilde{c}}{24}(\delta f'''(w)+\delta \bar{f}'''(\bar{w})).
    \label{perturbedenergystresstensordS}
\ea
Then, we can show
\ba
\begin{split}
   & 2\pi \int_{x_1}^{x_2} dx \frac{(\frac{x_2-x_1}{2})^2-(x-\frac{x_2+x_1}{2})^2}{x_2-x_1} T_{yy}  \\
    &\simeq -\frac{i \tilde{c}}{12} \int_0^{x_0} dx \frac{(\frac{x_2-x_1}{2})^2-(x-\frac{x_2+x_1}{2})^2}{x_2-x_1} (\delta f'''(w)+\delta \bar{f}'''(\bar{w})) \\
    &\simeq\frac{i\tilde{c}}{6}\Bigg(\frac{\delta f(w_2)+\delta \bar{f}(\bar{w}_2)-\delta f(w_1)-\delta \bar{f}(\bar{w}_1)}{x_2-x_1} -\frac{\delta f'(w_2)+\delta \bar{f}'(\bar{w}_2)+\delta f'(w_1)+\delta \bar{f}'(\bar{w}_1)}{2}\Bigg). \\
    \label{DeltaEA}
\end{split}
\ea
This result coincides with (\ref{eq:dSDeltaS}) and thus 
we confirmed the first law (\ref{eq:modHonplane}) in dS$_3$/CFT$_2$ correspondence. We can also formally derive the expression of entanglement entropy and first law from the Banados map in dS$_3/$CFT$_2$ as we explain in the appendix \ref{ap:Banados}. \\

\subsection{Real Space Ansatz}\label{sec:realg}
 In this section, we consider the global Lorentzian dS$_3$, which is created by a Euclidean instanton from the vacuum in a standard way. We would like to compute the variation of the entanglement entropy under the metric perturbation.
%\subsubsection{Geodesic}

  First, let us determine the profile of the geodesic for the computation of the entanglement entropy. The metric of a global dS$_3$ is given by 
  \be
  ds^2=-dt^2+\cosh^2t (d\theta^2+\sin^2\theta d\phi^2),
  \ee
  where 
  \be
0\leq t<\infty,\quad 0\leq\theta\leq\pi,\quad 0\leq\phi\leq2\pi.
  \ee
 According to the Hartle-Hawking prescription, it is natural to analytically continue the global dS$_3$ into the Euclidean signature by setting $t=i\tau$. The resulting geometry is given by a semi-sphere, whose metric reads
 \be
ds^2=d\tau^2+\cos^2\tau(d\theta^2+\sin^2\theta d\phi^2),
 \ee
 where 
 \be
-\frac{\pi}{2}\leq\tau\leq0,\quad 0\leq\theta\leq\pi,\quad 0\leq\phi\leq2\pi.
 \ee
  The subsystem $A$ in the CFT at the future infinity 
  is defined by
  \be\label{eq:subsystem_theta}
t=t_\infty,\quad \theta_1\leq\theta\leq\theta_2,\quad \phi=\phi_*.
  \ee
There are no smooth geodesics connecting these two end points if they pass through only  the Lorentzian geometry. However, by considering the Euclidean semi-sphere, it is possible to connect these end points as argued in \cite{Hikida:2022ltr}. To determine the profile, we compute the geodesic length. Suppose that  geodesics emanate from $\theta=\theta_1$ and $\theta_2$ and pass $\theta=\tilde{\theta}_1$ and $\tilde{\theta}_2$ at $t=0$, respectively. The sum of these geodesic lengths in the global dS$_3$ reads
\be
D_L=\arccos[\cosh t_\infty \cos(\theta_1-\tilde{\theta}_1)]+\arccos[\cosh t_\infty \cos(\theta_2-\tilde{\theta}_2)].
\ee
 For the Euclidean semi-sphere, the geodesic length connecting $\theta=\tilde{\theta}_1,\tilde{\theta}_2$ is simply given by
 \be
D_E=|\tilde{\theta}_2-\tilde{\theta}_1|.
 \ee
 In total, the length is expressed as  follows
 \be
D=D_L+D_E=\arccos[\cosh t_\infty \cos(\theta_1-\tilde{\theta}_1)]+\arccos[\cosh t_\infty \cos(\theta_2-\tilde{\theta}_2)]+|\tilde{\theta}_2-\tilde{\theta}_1|.
\ee
The variational problem considered in \cite{Hikida:2022ltr} is to maximize the real part of the geodesic distance, leading to 
\be
\tilde{\theta}_2-\tilde{\theta}_1=\pi,\quad \delta\tilde{\theta}_2=\delta\tilde{\theta}_1.
\ee
  By extremizing the imaginary part, we also obtain
  \be
\tilde{\theta}_1=\frac{\theta_1+\theta_2-\pi}{2},\quad\tilde{\theta}_2=\frac{\theta_1+\theta_2+\pi}{2}.
  \ee
Under this configuration, the Lorentzian profile is uniquely determined. By solving the Euler-Lagrange problem, the profiles are explicitly given by
\begin{align}
\tan\left[\theta-\frac{\theta_1+\theta_2-\pi}{2}\right]&=\frac{\sinh t}{\sqrt{1+\tan^2\frac{\theta_1-\theta_2}{2}\cosh^2 t}},\nn\\
\tan\left[\theta-\frac{\theta_1+\theta_2+\pi}{2}\right]&=-\frac{\sinh t}{\sqrt{1+\tan^2\frac{\theta_1-\theta_2}{2}\cosh^2 t}},
\end{align}
where the first (second) one passes through $\theta=\theta_1$ ($\theta_2$) at $t=t_\infty$.

However, the geodesic in the Euclidean semi-sphere is not determined since every geodesic passing through the great circle gives the same length. The geodesic in the Euclidean semi sphere is obtained directly
\begin{align}
\tan\left(\theta-\frac{\theta_1+\theta_2-\pi}{2}\right)&=-\frac{\sin\tau}{\sqrt{-1+\frac{\cos^2\tau}{\cos^2\tau_*}}},\\
\tan\left(\theta-\frac{\theta_1+\theta_2+\pi}{2}\right)&=\frac{\sin\tau}{\sqrt{-1+\frac{\cos^2\tau}{\cos^2\tau_*}}},
\end{align}
where the first one runs $\frac{\theta_1+\theta_2-\pi}{2}\leq\theta\leq\frac{\theta_1+\theta_2}{2}$ and the second one does $\frac{\theta_1+\theta_2}{2}\leq\theta\leq\frac{\theta_1+\theta_2+\pi}{2}$. The $\tau_*$ is the turning point of the geodesic. Unlike the Lorentzian case, the profile is not fully determined yet.

\subsection{Perturbations of dS$_3$}
To compute the change of holographic pseudo entropy, we analyze perturbations of the background geometry described by the metric:
\be
ds^2=-dt^2+\cosh^2 t~ g_{ij}dx^idx^j,
\ee
  where we set
  \be
g_{\theta\theta}=1+h_{\theta\theta}(t,\theta,\phi),\quad g_{\theta\phi}=h_{\theta\phi}(t,\theta,\phi),\quad g_{\phi\phi}=\sin^2\theta+h_{\phi\phi}(t,\theta,\phi).
  \ee
We impose the Einstein equation $R_{ab}-\frac{1}{2}Rg_{ab}+g_{ab}=0$ up to the first order in $h_{ij}$.  The $tt$ component leads to the traceless condition
  \be
h_{\phi\phi}=-\sin^2\theta h_{\theta\theta}.
  \ee
  
The $ij$ components of Einstein equation lead to
\ba
2\tanh t \de_t h_{ij}+\de_t^2 h_{ij}=0,
\ea
which is solved as \ba
h_{ij}\propto (1-\tanh t). \label{wwwab}
\ea
Here, we impose a boundary condition
$\lim_{t\to\infty}h_{ij}=0$, which means that we do not change the boundary metric. The remaining equations are given by
\begin{align}
&\left(\de^2_\phi+\sin^2\theta\de_\theta^2\right)H+5\sin\theta\cos\theta \de_\theta H+(4\cos^2\theta-2\sin^2\theta)H=0, \nn\\
&\left(\de^2_\phi+\sin^2\theta\de_\theta^2\right)J+3\sin\theta\cos\theta \de_\theta J+(2\cos^2\theta-1)J=0,\label{eomwq}
\end{align}
where  we introduced $H(\theta,\phi)$ and $J(\theta,\phi)$ by
\begin{align}
  & h_{\theta\theta}=(1-\tanh t)\cdot H(\theta,\phi),\nn\\
&h_{\phi\phi}=-(1-\tanh t)\cdot \sin^2\theta\cdot H(\theta,\phi), \nn\\
& h_{\theta\phi}=(1-\tanh t)\cdot J(\theta,\phi),\nn\\
& \de_\theta[\sin\theta J(\theta,\phi)]=\sin\theta\de_\phi H(\theta,\phi),
\nn\\
&\de_\phi J(\theta,\phi)=-\de_\theta\left[\sin^2\theta H(\theta,\phi)\right].
\label{metricsolution}
\end{align}
The analytical solutions are given by
\begin{align}
& H(\theta,\phi)=H(\theta)e^{im\phi},\quad  
J(\theta,\phi)=J(\theta)e^{im\phi}, 
\nn\\
& H(\theta)=\frac{1}{(\sin\theta)^2}\left[A_1\left(\tan\frac{\theta}{2}\right)^m
+A_2\frac{1}{\left(\tan\frac{\theta}{2}\right)^{m}}\right],\nn\\
& J(\theta)=\frac{i}{\sin\theta}\left[A_1\left(\tan\frac{\theta}{2}\right)^m
-A_2\frac{1}{\left(\tan\frac{\theta}{2}\right)^{m}}\right],
\label{HJm}
\end{align}
where $A_1$ and $A_2$ are integration constants. 

The holographic energy stress tensor for the dual CFT$_2$ can be found from the perturbation $h_{ij}$ in the limit $t\to\infty$ as in the familiar case of AdS/CFT \cite{Balasubramanian:1999re,deHaro:2000vlm}. In our dS$_3/$CFT$_2$, this reads\footnote{The imaginary factor arises from the analytic continuation of the radius.}
\ba
T_{ij}=\frac{i}{32\pi G_N}\lim_{t\to\infty}e^{-2t}h_{ij}(t,\theta,\phi). 
\label{EMDS}
\ea
This is governed by the functions $H$ and $J$. One may note that there are singularities at $\theta=0$ and $\theta=\pi$ for a mode with a fixed value of $m$. This is interpreted as a class of excited states, which are created by inserting primary or descendant operators at the two points as we explain in appendix  
\ref{ap:EM}.

In the Euclidean case, we can obtain the solution by the analytic continuation $\tau=it$ from the Lorentzian one, which leads to
 \be
ds^2=d\tau^2+\cos^2\tau g_{ij}dx^i dx^j,
 \ee
  where 
    \be
g_{\theta\theta}=1+h_{\theta\theta}(\tau,\theta,\phi),\quad g_{\theta\phi}=h_{\theta\phi}(\tau,\theta,\phi),\quad g_{\phi\phi}=\sin^2\theta+h_{\phi\phi}(\tau,\theta,\phi),
  \ee
   and 
   \begin{align}
  & h_{\theta\theta}=(1-i
  \tan \tau)\cdot H(\theta,\phi),\nn\\
&h_{\phi\phi}=-(1-i\tan \tau)\cdot \sin^2\theta\cdot H(\theta,\phi), \nn\\
& h_{\theta\phi}=(1-i\tan \tau)\cdot J(\theta,\phi),
\end{align}
where $H(\theta,\phi)$ and $J(\theta,\phi)$ are the same as the previous ones.

\subsection{Computation of $\Delta S_A$ for $m=0$}
As in the AdS/CFT case, the change $\Delta S_A$ of holographic pseudo entropy can be calculated from (\ref{CHEE}). For simplicity, consider the $m=0$ mode of the perturbation in (\ref{HJm}). The metric with the perturbation reads ($\lambda$ is an infinitesimal deformation parameter):
\ba
ds^2&=&-dt^2+\cosh^2t\Biggl[\left(1+\lambda\frac{1-\tanh t}{\sin^2\theta}\right)d\theta^2+2i\lambda
\frac{(1-\tanh t)}{\sin\theta}d\theta d\phi\no
&&+\left(\sin^2\theta-\lambda(1-\tanh t)\right)d\phi^2\Biggr].
\ea
The geodesic in the computation of $\Delta S_A$ is on the time slice as $\phi=\phi_*$. The induced metric reads
\be
ds^2=\left[-1+\left(\cosh^2 t+\lambda\frac{\cosh^2 t(1-\tanh t)}{\sin^2\theta}\right)\theta'^2\right] dt^2,
\ee
where $\theta'=\frac{d\theta}{d t}$.  Substituting this for the Lorentzian part,  we obtain 
\begin{align}
    \Delta S_{\rm L}&=\frac{\l}{8G}\int_{0}^{t_\infty}dt\frac{\cosh^2t (1-\tanh t)\theta'^2}{\sqrt{-1+\cosh^2t \theta'^2}\sin^2\theta}\nn\\
%    &=-\frac{i}{8G}\int_{0}^{t_\infty}dt\frac{ (1-\tanh t)}{\tan \frac{\theta_2-%\theta_1}{2}\cosh t \sqrt{1+\tan^2 \frac{\theta_2-\theta_1}{2}\cosh^2 t} }\nn\\
%    &\cdot\left(\frac{1}{\left(-\frac{1-\tanh^2 t \cos^2\frac{\theta_2-\theta_1}
%{2}}{\cosh t }\cos\frac{\theta_1+\theta_2}{2}+\tanh t \cos\frac{\theta_2-\theta_1}
%{2}\sin\frac{\theta_1+\theta_2}{2}\right)^2}\right.\nn\\
%   &+ \left.\frac{1}{\left(\frac{1-\tanh^2 t \cos^2\frac{\theta_2-\theta_1}{2}}
%{\cosh t }\cos\frac{\theta_1+\theta_2}{2}+\tanh t \cos\frac{\theta_2-\theta_1}
%{2}\sin\frac{\theta_1+\theta_2}{2}\right)^2}\right)\nn\\
   &=-\frac{i\lambda}{4G}\left(\frac{\sin\frac{\theta_1+\theta_2}{2}}{\sin\frac{\theta_2-\theta_1}{2}}\arctanh\left[\frac{1}{\sin\frac{\theta_1+\theta_2}{2}}\right]+\frac{1}{\sin\frac{\theta_2-\theta_1}{2}}\right.
   \left.+\frac{\sin\frac{\theta_1+\theta_2}{2}}{\sin\frac{\theta_2-\theta_1}{2}}\arctanh\left[\frac{\sin\frac{\theta_2-\theta_1}{2}}{\sin\frac{\theta_1+\theta_2}{2}}\right]-1\right).
\end{align}

For the Euclidean part, we find
\begin{align}
    \Delta S_{\rm E}&=\frac{\lambda}{8G}\int_{\tau_*}^0d\tau \frac{(\cos^2\tau-i\cos\tau\sin\tau)\theta'^2}{\sqrt{1+\cos^2\tau\theta'^2}\sin^2\theta} \nn\\
%    &=\frac{1}{8G}\int_{\tau_*}^0d\tau \frac{\cos^2\tau_*(1-i\tan\tau)}
%{\cos\tau\sqrt{\cos^2\tau-\cos^2\tau_*}}\nn\\
%    &\cdot\left(\frac{1}{\left(\frac{\tan\tau}
%{\tan\tau_*}\sin\frac{\theta_1+\theta_2}{2}-\sqrt{1-\frac{\tan^2\tau}
%{\tan^2\tau_*}}\cos\frac{\theta_1+\theta_2}{2}\right)^2} +\frac{1}
%{\left(\frac{\tan\tau}{\tan\tau_*}\sin\frac{\theta_1+\theta_2}{2}+\sqrt{1-%\frac{\tan^2\tau}{\tan^2\tau_*}}\cos\frac{\theta_1+\theta_2}
%{2}\right)^2}\right)\nn\\
    &=\frac{i\lambda\sin\tau_*}{4G}\left(-\sin\frac{\theta_1+\theta_2}{2}\arctanh\left[\frac{1}{\sin\frac{\theta_1+\theta_2}{2}}\right]+1\right),
\end{align}
where $-\frac{\pi}{2}\leq\tau_*\leq0$ is the turning point of the geodesic in the semi-sphere.
In total, we obtain
\begin{align}
    \Delta S=\frac{i\lambda}{4G}&\left(\left(\sin\frac{\theta_1+\theta_2}{2}\arctanh\left[\frac{1}{\sin\frac{\theta_1+\theta_2}{2}}\right]-1\right)\left(\frac{1}{\sin\frac{\theta_2-\theta_1}{2}}-\sin\tau_*\right)\right.\nn\\
    &\left.-\frac{\sin\frac{\theta_1+\theta_2}{2}}{\sin\frac{\theta_2-\theta_1}{2}}\arctanh\left[\frac{\sin\frac{\theta_2-\theta_1}{2}}{\sin\frac{\theta_1+\theta_2}{2}}\right]+1\right).
\end{align}
The second term corresponds to that which reproduces the first law (\ref{modHonsphere}), and thus if the first term vanishes, the first law holds\footnote{The term that correctly reproduces the first law comes from the upper limit of the primitive function. This signals that the term regarding the cutoff successfully satisfies the first law as in the AdS case. In other words, the contributions from  the lower limit or $t=0$ breaks the first law. This is probably because the geodesics are not smoothly connected at $t=0$, which makes the variational formula of the length ill-defined.   }. This is only true when the $\tau_*$ takes a complex value\footnote{For the case with $\theta_2-\theta_1=\pi$, the first law holds if $\tau_*=\frac{\pi}{2}$. However, this is out of the domain. }, which represents that the background geometry is not a semi-sphere anymore. In this way, the identification of the geodesic $\Gamma_A$ in (\ref{RT}) for the holographic entanglement entropy with the union of the space-like geodesic and time-like one 
presented in subsection \ref{sec:realg} no longer works in the presence of perturbations. This urges us to introduce geodesics in a complexified de Sitter space as we will study in the next section.

%%%%%%%%%%%%%%%%%%%%%%%%%%%%%%%%%%%%%%%%%%%%%%%%%%%%%%%%
%%%%%%%%%%%%%%%%%%%%%%%%%%%%%%%%%%%%%%%%%%%%%%%%%%%%%%%%
\section{First law from complex geometry in dS/CFT}
\label{sec:dsCFTF}
%%%%%%%%%%%%%%%%%%%%%%%%%%%%%%%%%%%%%%%%%%%%%%%%%%%%%%%%
%%%%%%%%%%%%%%%%%%%%%%%%%%%%%%%%%%%%%%%%%%%%%%%%%%%%%%%%

Our analysis in the previous section shows that we cannot find the correct geodesic $\Gamma_A$ in (\ref{RT}) within the real space of dS$_3$.
Thus, in this section, we would like to explore the geodesic in the complexified coordinates of dS$_3$,  which reproduces the expected holographic pseudo entropy. 
For this, we take a close look at the holographic entanglement entropy in AdS$_3$ with infinitesimally small perturbations around the pure AdS$_3$ and perform its analytical continuation to dS$_3$. This strategy is very similar to the systematic calculation method pioneered in \cite{Heller:2024whi,Heller:2025kvp} for the computation of time-like entanglement entropy \cite{Doi:2022iyj,Doi:2023zaf}.

Considering the case where the asymptotic boundary is given by a sphere S$^2$,  the perturbed metric of the Euclidean AdS$_3$ and the Lorentzian dS$_3$ metric is expressed as
\ba
 \mbox{AdS$_3$}&:&ds^2=d\rho^2+\sinh^2\rho (d\theta^2+\sin^2\theta d\phi^2+\ti{h}_{ij}dx^i dx^j),\label{pAdS} \\
\mbox{dS$_3$}&:&ds^2=-dt^2+\cosh^2 t (d\theta^2+\sin^2\theta d\phi^2+h_{ij}dx^i dx^j).
\ea
They are related with each other by the coordinate transformation
\ba
t=\rho-\frac{\pi}{2}i,
\ea
with the AdS radius replaced by $i$ times the dS radius. Here, we employed the global dS$_3$ again instead of Poincar\'{e} dS$_3$. This is because it is more convenient and clearer than other coordinates when we consider the continuation to the Euclidean instanton. Nevertheless, we can apply the arguments to other coordinate systems. In particular, we presented the analysis for the Poincar\'{e} dS$_3$ in appendix \ref{ap:poin}. 

Since we have the global dS$_3$ in mind, first we need to examine the holographic entanglement entropy in the three dimensional hyperbolic ball (\ref{pAdS}) below. To see the relation of the holographic (pseudo) entropy in AdS$_3$ and dS$_3$, it is useful to rewrite the expression of the geodesic length in terms of the integral over $\theta$ as we will see below.

%In this proposal, the holographic time-like entanglement entropy (HTEE) is %associated with codimension-two extremal surfaces $\gamma_T$, which are anchored %on a timelike boundary subregion $T$ and generally extend into a complexified %bulk geometry. 
%The HTEE is then defined as being proportional to the area of $\gamma_T$, 
%\begin{align}
%    S_T = \frac{A(\gamma_T)}{4G_N},
%\end{align}
%where $G$ denotes the bulk gravitational constant.
% The normalization is chosen such that the standard HEE is recovered upon %analytic continuation.

\subsection{Holographic entanglement on a 3d hyperbolic ball}

Before the perturbation $\ti{h}$ was introduced in (\ref{pAdS}), the geodesic which connects two points $(\theta,\phi)=(\theta_1,0)$ and $(\theta_2,0)$  is specified by the following functional relation between $\rho$ and $\theta$:
\ba
&&\sinh \rho=\frac{1}{\s{\sin^2\left(\theta+\frac{\pi}{2}-\frac{\theta_1+\theta_2}{2}\right)\tan^2\frac{\theta_2-\theta_1}{2}-\cos^2\left(\theta+\frac{\pi}{2}-\frac{\theta_1+\theta_2}{2}\right)}},\no
%&&\cosh \rho=\frac{\sin\left(\theta+\frac{\pi}{2}-\frac{\theta_1+\theta_2}
%{2}\right)}{\s{\sin^2\left(\theta+\frac{\pi}{2}-\frac{\theta_1+\theta_2}
%{2}\right)\sin^2\frac{\theta_2-\theta_1}{2}-\cos^2\left(\theta+\frac{\pi}{2}-%\frac{\theta_1+\theta_2}{2}\right)\cos^2\frac{\theta_2-\theta_1}{2}}}.
\ea

\begin{figure}[hhh]
		\centering
		\includegraphics[width=12cm]{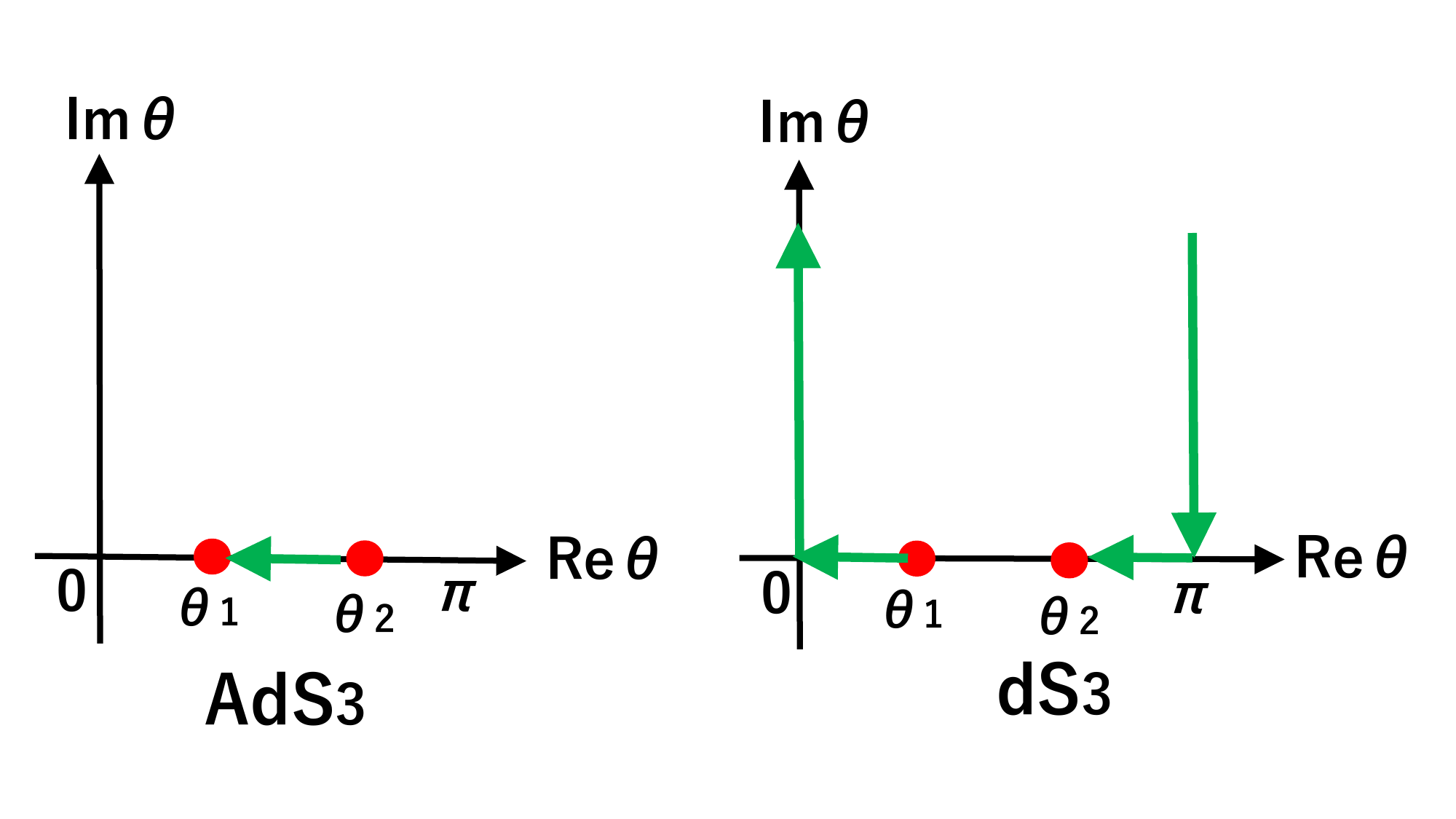}
		\caption{Contours of $\theta$ integral of the entropy calculation in the AdS$_3$ (left) and dS$_3$ (right). In the above picture we took $\theta_1+\theta_2=\pi$ and thus $\psi_1=0$ and $\psi_2=\pi$.} 
		\label{fig:contour}
\end{figure}

Its geodesic length reads
\ba
D_{AdS}&=&\int^{\theta_2-\ep}_{\theta_1+\ep} d\theta \s{\dot{\rho}^2+\sinh^2\rho}\no
&=&\int^{\theta_2-\ep}_{\theta_1+\ep} d\theta \frac{\tan\frac{\theta_2-\theta_1}{2}}{\sin^2\left(\theta+\frac{\pi}{2}-\frac{\theta_1+\theta_2}{2}\right)\tan^2\frac{\theta_2-\theta_1}{2}-\cos^2\left(\theta+\frac{\pi}{2}-\frac{\theta_1+\theta_2}{2}\right)},
\ea
where the integration contour is the obvious one depicted in the left panel of Fig.\ref{fig:contour}. We chose the UV cut off as 
\ba
\ep=\frac{2e^{-2\rho_\infty}}{\tan\frac{\theta_2-\theta_1}{2}}.
\ea
This is explicitly evaluated as
\ba
D_{AdS}=2\rho_\infty+\log \sin^2\frac{\theta_2-\theta_1}{2}.
\ea

Now, we would like to calculate the change of the geodesic length due to the perturbation. By using (\ref{CHEE}), this is computed as follows
\ba
\Delta D_{AdS}&=&\frac{1}{2}\int^{\theta_2}_{\theta_1} d\theta \frac{1}{\tan\frac{\theta_2-\theta_1}{2}}\left(1-\frac{\cosh\rho}{\sinh\rho}\right)H(\theta,\phi)\no
&=&\frac{1}{2}\int^{\theta_2}_{\theta_1} d\theta \frac{1}{\tan\frac{\theta_2-\theta_1}{2}}\left(1-\frac{\sin\left(\theta+\frac{\pi}{2}-\frac{\theta_1+\theta_2}{2}\right)}{\cos\frac{\theta_2-\theta_1}{2}}\right)H(\theta,\phi),
\label{AdShee}
\ea
where the integration contour is the obvious one depicted in the left panel of Fig.\ref{fig:contour}.

We introduced the holographic energy stress tensor $T^{AdS}_{ij}$ in AdS$_3/$CFT$_2$ \cite{Balasubramanian:1999re,deHaro:2000vlm}, given by
\ba
T^{AdS}_{ij}=-\frac{1}{32\pi G_N}\lim_{\rho\to\infty}e^{-2\rho}\ti{h}_{ij}(\rho,\theta,\phi). 
\ea
Since in the AdS boundary limit $\rho\to \infty$, we have
$\ti{h}_{\phi\phi}\simeq 2e^{-2\rho} \sin^2\theta H(\theta,\phi)$,
 leading to 
\ba
T^{AdS}_{\phi\phi}=-\frac{1}{16\pi G_N}H(\theta,\phi).
\ea
Thus, we can rewrite (\ref{AdShee}) as
\ba
\Delta D_{AdS}&=&-8\pi G_N\int^{\theta_2}_{\theta_1} d\theta \frac{1}{\tan\frac{\theta_2-\theta_1}{2}}\left(1-\frac{\sin\left(\theta+\frac{\pi}{2}-\frac{\theta_1+\theta_2}{2}\right)}{\cos\frac{\theta_2-\theta_1}{2}}\right)\frac{T^{AdS}_{\phi\phi}}{\sin^2\theta},\no
&=&8\pi G_N\int^{\theta_2}_{\theta_1}d\theta \frac{\tan^2\frac{\theta}{2}}{2\sin^2\left(\frac{\theta}{2}\right)}\cdot \frac{\sin\left(\frac{\theta-\theta_1}{2}\right)\sin\left(\frac{\theta_2-\theta}{2}\right)}{\sin^2\left(\frac{\theta}{2}\right)\sin\left(\frac{\theta_2-\theta_1}{2}\right)}   T^{AdS}_{\phi\phi}.
\ea
Therefore,  the shift of the holographic entanglement entropy $\Delta S_A=\frac{\Delta D_{AdS}}{4G_N}$ reproduces the expected first law (\ref{modHonsphere}).

We can compute $\Delta S_A$ for each mode $m$ perturbation. We explicitly have 
\begin{align}
    \Delta S_A&=\frac{1}{4G_N}\int_{\theta_1}^{\theta_2}\frac{1}{\tan\frac{\theta_2-\theta_1}{2}}\left(1-\frac{\cos\left(\theta-\frac{\theta_1+\theta_2}{2}\right)}{\cos\left(\frac{\theta_2-\theta_1}{2}\right)}\right) \frac{\left(\tan\frac{\theta}{2}\right)^m}{\sin^2\theta}e^{im\phi}\nn\\
    &=-\frac{1}{4G}\frac{\left(m+\frac{\sin b}{\sin a}\right)\left(\tan\frac{b-a}{2}\right)^m+\left(m-\frac{\sin b}{\sin a}\right)\left(\tan\frac{b+a}{2}\right)^m}{m(m^2-1)}e^{im\phi},
\end{align}
where $a=\frac{\theta_2-\theta_1}{2}$ and $b=\frac{\theta_2+\theta_1}{2}$.
Notice that this $\Delta S$ satisfies the following differential equation
\be
\left[\frac{\de}{\de\theta_1}\frac{\de}{\de\theta_2}+\frac{2}{\sin^2\left(\frac{\theta_2-\theta_1}{2}\right)}\right]\Delta S_A=0.
\ee
Indeed, this can be rewritten as $(\Box_{dS_2}+2)\Delta S_A=0$ with the dS$_2$ metric $ds^2=\frac{d\theta_1 d\theta_2}{\sin^2\left(\frac{\theta_2-\theta_1}{2}\right)}$, generalizing the result (\ref{FSEOM}) for the Poincar\'{e} coordinate to this hyperbolic coordinate.

%%%%%%%%%%%%%%%%
\subsection{Holographic pseudo entropy from complex geodesics in dS$_3$}
%%%%%%%%%%%%%%%%%%%%%

Now, we  would like to derive the first law from the bulk side, allowing for possible extensions into a complexified geometry. 
We choose the boundary subsystem as in \eqref{eq:subsystem_theta}.
Then the geodesic connecting $(\theta_1,0)$ and $(\theta_2,0)$ is given by  
\ba
\cosh t &=& 
\frac{1}{\sqrt{-\sin^2\!\left(\theta+\frac{\pi}{2}-\frac{\theta_1+\theta_2}{2}\right)\tan^2\frac{\theta_2-\theta_1}{2}
+\cos^2\!\left(\theta+\frac{\pi}{2}-\frac{\theta_1+\theta_2}{2}\right)}}, 
\label{geods}
\ea
The leading geodesic length (before including the perturbation $h_{ij}$) takes the form  
\ba
L_\text{dS} &=& i \int^{\theta_2+\delta}_{\theta_1-\delta} d\theta \, \sqrt{\dot{t}^{\,2}-\cosh^2 t} \no \\
&=& i \int^{\theta_2+\delta}_{\theta_1-\delta} d\theta \,\notag 
\frac{\tan \tfrac{\theta_2-\theta_1}{2}}
{-\sin^2\!\left(\theta+\tfrac{\pi}{2}-\tfrac{\theta_1+\theta_2}{2}\right)\tan^2\tfrac{\theta_2-\theta_1}{2}
+\cos^2\!\left(\theta+\tfrac{\pi}{2}-\tfrac{\theta_1+\theta_2}{2}\right)}.
\ea
Here, we chose the contour by a natural deformation of the previous one (shown in the left panel of Fig.~\ref{fig:contour}), namely the contour shown in the right panel of Fig.~\ref{fig:contour}, which extends into the complexified geometry.\footnote{More explicitly, the contour is defined as follows. It starts from $(t,\theta)=(t_\infty,\theta_1)$ and proceeds to $(t,\theta)=\left(0,-\tfrac{\theta_2-\theta_1}{2}+\tfrac{\pi}{2}\right)$ in Lorentzian dS$_3$. 
It then enters Euclidean dS$_3$ by setting 
$\theta=i\eta-\tfrac{\pi}{2}+\tfrac{\theta_1+\theta_2}{2}$ and $t=i\tau$, 
with $\eta,\tau \in \mathbb{R}$ satisfying  
\ba
\cos \tau = \frac{1}{\sqrt{\sinh^2 \eta \, \tan^2 \tfrac{\theta_2-\theta_1}{2} + \cosh^2 \eta}},
\ea
thus following the geodesic (\ref{geods}). The contour reaches $\eta \to \infty$ and $\tau=\tfrac{\pi}{2}$, then turns around by taking 
$\theta=i\tilde{\eta}+\tfrac{\pi}{2}+\tfrac{\theta_1+\theta_2}{2}$ with $\tilde{\eta}$ real, and finally ends at the other dS boundary point $(t_\infty,\theta_2)$. 
This contour corresponds to the one shown in the right panel of Fig.~\ref{fig:contour}, and is analogous to the construction in \cite{Heller:2024whi} for the time-like entanglement entropy.} Though this choice of the contour coincides with the one used in subsection \ref{sec:realg} for the Lorentzian part of dS$_3$, they deviate for the Euclidean part. The previous one is a real geodesic in the Euclidean dS$_3$ (semi-sphere), while the present one extends into a complexified direction. Such a new choice of the geodesic is crucial to obtain the correct result which reproduces the first law as we will see below.

Evaluating the integral\footnote{
The UV cutoff $\delta$ is chosen with the opposite sign compared to AdS, i.e., $\delta=-\epsilon$, which is explicitly given by  
\ba
\delta = \frac{2e^{-2t_\infty}}{\tan \tfrac{\theta_2-\theta_1}{2}}.
\ea
}, we obtain  
\ba
L_\text{dS} = 2 i t_\infty + i \log \sin^2 \tfrac{\theta_2-\theta_1}{2} + \pi,
\ea
which reproduces the known expression \cite{Doi:2022iyj,Doi:2023zaf} of holographic pseudo entropy via the formula (\ref{RT}):
\begin{align}
    S_{A}=i\frac{\ti{c}}{6}\log \left[\frac{1}{\ep^2} \sin^2 \tfrac{\theta_2-\theta_1}{2}\right]+\frac{\ti{\pi c}}{6}.
\end{align}

We perform the expansion of $h_{\phi\phi}$ in the near dS boundary limit $t\rightarrow\infty$
\ba
    h_{\phi\phi}\simeq -2e^{-2t}\sin^2\theta H(\theta,\phi) .
\ea
By applying the expression (\ref{EMDS}) of the holographic stress tensor in dS$_3/$CFT$_2$, we obtain
\ba
T_{\phi\phi}=&&\frac{-i}{16\pi G_N}\sin^2\theta H(\theta,\phi). \label{T_phiphi}
\ea
The variation of the geodesic length due to the perturbation is computed as  
\ba
\Delta D_{dS}=&&\frac{1}{2}\int^{\theta_2}_{\theta_1}     
    d\theta \frac{i}{\tan\frac{\theta_2-\theta_1}{2}}\left(1-\frac{\cos(\theta-\frac{\theta_2+\theta_1}{2})}{\cos\frac{\theta_2-\theta_1}{2}}\right) H(\theta,\phi) \\
=&&-8\pi G_N\int^{\theta_2}_{\theta_1} d\theta \frac{1}{\tan\frac{\theta_2-\theta_1}{2}}\left(1-\frac{\sin\left(\theta+\frac{\pi}{2}-\frac{\theta_1+\theta_2}{2}\right)}{\cos\frac{\theta_2-\theta_1}{2}}\right)\frac{T_{\phi\phi}}{\sin^2\theta}
\ea
for a generic profile of $T_{\phi\phi}$. \\
% We take the time slice $\phi=0$, equivalently $\tau=0$, using the conformal map 
% \ba
% x + i \tau = e^{i\phi} \cot \frac{\theta}{2}.
% \ea
% On the plane at $\tau=0$, the first law in 2d CFT reads
% \ba
% H_A = \int^{x_2}_{x_1} dx \, \frac{l^2 - |x - \xi|^2}{2l} \, T_{\tau\tau},
% \ea
% where 
% \ba
% x_1 = \cot \frac{\theta_1}{2}, \quad 
% x_2 = \cot \frac{\theta_2}{2}, \quad
% l = \frac{x_2 - x_1}{2}, \quad 
% \xi = \frac{x_1 + x_2}{2}.
% \ea
% A direct computation shows that
% \begin{align}
%     &\frac{l^2 - |x - \xi|^2}{2l} 
% = \frac{\sin\!\left(\tfrac{\theta-\theta_1}{2}\right) \sin\!\left(\tfrac{\theta_2-\theta}{2}\right)}
% {\sin^2 \!\left(\tfrac{\theta}{2}\right) \sin\!\left(\tfrac{\theta_2-\theta_1}{2}\right)},\\
% &T_{\tau\tau} 
% = \left(\frac{d\phi}{d\tau}\right)^2 T_{\phi\phi} 
% = \left(\tan \tfrac{\theta}{2}\right)^2 T_{\phi\phi}.
% \end{align}
Finally, the change of the holographic pseudo entropy is evaluated to be 
\ba
\Delta S_A &=& \frac{\Delta D_{dS}}{4G_N}=
2\pi\int^{\theta_2}_{\theta_1}d\theta \frac{\tan^2\frac{\theta}{2}}{2\sin^2\left(\frac{\theta}{2}\right)}\cdot \frac{\sin\left(\frac{\theta-\theta_1}{2}\right)\sin\left(\frac{\theta_2-\theta}{2}\right)}{\sin^2\left(\frac{\theta}{2}\right)\sin\left(\frac{\theta_2-\theta_1}{2}\right)}   T_{\phi\phi}. \notag \\
%&=& 
%-2\pi\int^{\theta_2}_{\theta_1} d\theta \frac{1}{\tan\frac{\theta_2-\theta_1}
%{2}}\left(1-\frac{\sin\left(\theta+\frac{\pi}{2}-\frac{\theta_1+\theta_2}
%{2}\right)}{\cos\frac{\theta_2-\theta_1}{2}}\right)\frac{T_{\phi\phi}}
%{\sin^2\theta}
\label{deSdss}
\ea
This successfully reproduces the first law of pseudo entropy (\ref{modHonsphere}) as promised. \\
Here, we would like to notice that we can choose any complex contour which connects $\theta_1$ and $\theta_2$, as long as it avoids the poles at $\theta=0,\pi$. In the right panel of Fig. \ref{fig:contour}, we show  that the contour that passes through the real axis, following the previous choice of the geodesic for Lorentzian dS$_3$.

% Finally, by introducing 
% \ba
% \Delta \theta = \frac{\theta_2 - \theta_1}{2}, 
% \qquad 
% \zeta = \frac{\theta_1 + \theta_2}{2},
% \ea
% we can rewrite the integral by shifting $\theta \to \theta + \zeta$: 
% \ba
% \Delta L_\text{dS}(\zeta,\Delta \theta,\phi) &=& 
% i \int^{\Delta \theta}_{-\Delta \theta} d\theta \,
% \frac{1}{\tan \Delta \theta}
% \left(\frac{\cos \theta}{\cos \Delta \theta} - 1\right) 
% H(\theta+\zeta,\phi).
% \ea

%%%%%%%%%%%%%%%%%%%%%%%%%%%%%%%%%%%%%%%%%%%%%%%%%%%%%%%%
%%%%%%%%%%%%%%%%%%%%%%%%%%%%%%%%%%%%%%%%%%%%%%%%%%%%%%%%
\section{Conclusions and Discussions}
\label{sec:conclusions}
%%%%%%%%%%%%%%%%%%%%%%%%%%%%%%%%%%%%%%%%%%%%%%%%%%%%%%%%
%%%%%%%%%%%%%%%%%%%%%%%%%%%%%%%%%%%%%%%%%%%%%%%%%%%%%%%%
In AdS/CFT, the entanglement entropy in a CFT is given by the area of an extremal surface \cite{Ryu:2006bv,Ryu:2006ef,Hubeny:2007xt} and thus we expect that the gravitational spacetime emerges from quantum entanglement. One basic relation which bridges the dynamics of quantum entanglement and that of gravity is the first law of entanglement entropy \cite{Blanco:2013joa,Wong:2013gua,Bhattacharya:2012mi}, which is known to be equivalent to the perturbative Einstein equation in AdS \cite{Lashkari:2013koa,Faulkner:2013ica}. 

Motivated by the above success and by the phenomenological importance of de Sitter spaces, the main purpose of this paper is to extend this relation to the dS/CFT correspondence. In the dS/CFT, the entropy is known to take complex values \cite{Sato:2015tta,Narayan:2015vda,Doi:2022iyj,Doi:2023zaf} and thus it should be regarded as pseudo entropy \cite{Nakata:2021ubr}, more correctly. In this paper, we computed the change of holographic pseudo entropy under infinitesimally small perturbations in the dS$_3$ setup dual to excited states in CFT$_2$ and see if it agrees with the first law of pseudo entropy \cite{Mollabashi:2021xsd}.

In an earlier work \cite{Doi:2022iyj,Doi:2023zaf}, for the vacuum dS$_3$, the holographic calculation of pseudo entropy was performed and shown to reproduce the dual CFT result. This geodesic extends in the Lorentzian and Euclidean de Sitter space so that the coordinate on the geodesic is real-valued as usual. 
Thus, our first attempt in this paper was to take this real space ansatz of geodesic and compute the holographic pseudo entropy as the length of slightly deformed geodesic due to the metric perturbation. However, it turned out that this does not work as the resulting geodesic length does not agree with the first law.

Instead, as an alternative approach, we looked back to the AdS$_3$ calculations for a perturbation of a hyperbolic ball solution, for which the first law relation should hold. Then we transform this AdS$_3$ calculation into that for the dS$_3$ by a standard analytical continuation. This naturally derives the profile of geodesic in dS$_3$ and guarantees that it satisfies the first law. Interestingly, the resulting geodesic turned out to extend in the complex valued direction of dS$_3$, though it passes through real values of the coordinate for the Lorentzian dS$_3$ part. This implies that the calculation of holographic pseudo entropy requires a fully complexification of the original de Sitter spacetime, not only just replacing the initial part with an Euclidean instanton. 
This is very similar to the prescription considered in \cite{Heller:2024whi} for the time-like entanglement entropy \cite{Doi:2022iyj,Doi:2023zaf}.

We would also like to point out one more possibility for the computation of holographic pseudo entropy. Even if we start from the real space ansatz of geodesic for the vacuum dS$_3$ explained in section \ref{sec:realg}, we can perform a coordinate transformation in dS$_3$, which is analogous to the Ba\~{n}ados map \cite{Banados:1998gg,Roberts:2012aq} for the AdS$_3/$CFT$_2$, to derive the gravity dual for excited states in CFT$_2$ as we explained in the appendix \ref{ap:Banados}.  This transformation maps the profile of geodesic for the vacuum state to those for excited states and clearly this keeps the geodesic length up to the choice of UV cut off. Thus this choice of the geodesic also seems to work well for the dS$_3/$CFT$_2$ case, though the relation to the holographic entanglement entropy in AdS$_3/$CFT$_2$ is not clear. It would be intriguing to work out what is the correct prescription by examining the higher dimensional de Sitter spaces or the effects of matter fields in dS$_3$ in future works. It would be also important to consider the reconstruction of metric for the dS/CFT from the data of holographic pseudo entropy for a given excited state. 

Finally, we found that the infinitesimally small change of pseudo entropy $\Delta S_A$ in dS$_3
/$CFT$_2$, satisfies the scalar field equation of motion in dS$_2$, which is an extension of the known result \cite{Nozaki:2013vta,deBoer:2015kda,deBoer:2016pqk} in AdS$_3/$CFT$_2$. This emergent dS$_2$ can be qualitatively understood as a time-like slice of dS$_3$. Thus the first law of pseudo entropy for the CFT$_2$ implies the emergence of the time coordinate of its bulk dual namely dS$_3$. Thus this looks like a hint to understand the basic mechanism behind the emergence of time in dS/CFT.
This may suggest that the quantum circuit or tensor network \cite{Swingle:2009bg,Pastawski:2015qua,Hayden:2016cfa} which creates the quantum state dual to the CFT$_2$ corresponds to the dS$_2$ slice as similar to the argument in \cite{Milsted:2018yur,Takayanagi:2018pml}. This may be possible because the CFT$_2$ is non-unitary and the original Euclidean path-integral which creates the state in the Euclidean CFT$_2$ actually looks like Lorentzian as the Hamiltonian becomes anti-Hermitian. The path-integral optimization \cite{Caputa:2017urj,Caputa:2017yrh,Czech:2017ryf,Boruch:2020wax,Boruch:2021hqs} may provide a useful framework on this.
It would be very important to understand this better in future works.

%%%%%%%%%%%%%%%%%%%%%%%%%%%%%%%%%%%%%%%%%%%%%%%%%%%%%%%%
%%%%%%%%%%%%%%%%%%%%%%%%%%%%%%%%%%%%%%%%%%%%%%%%%%%%%%%%
\section*{Acknowledgements}

%%%%%%%%%%%%%%%%%%%%%%%%%%%%%%%%%%%%%%%%%%%%%%%%%%%%%%%%
%%%%%%%%%%%%%%%%%%%%%%%%%%%%%%%%%%%%%%%%%%%%%%%%%%%%%%%%

We are very grateful to Ori Fabio and Michal Heller for useful discussions. This work is supported by MEXT KAKENHI Grant-in-Aid for Transformative Research Areas (A) through the ``Extreme Universe'' collaboration: Grant Number 21H05187.
YS is supported by Grant-in-Aid for JSPS Fellows No.\ 23KJ1337.
 TT is also supported by Inamori Research Institute for Science and by JSPS Grant-in-Aid for Scientific Research (B) No.~25K01000. 
\appendix

\section{Analysis using Ba\~{n}ados-like map on dS$_3$}\label{ap:Banados}
In this appendix, we derive the first law using the Poincar\'{e} coordinate via Ba\~{n}ados-like map on dS$_3$. The Ba\~{n}ados map was originally constructed for an AdS$_3$ and gives a gravity dual of conformal maps in 2d CFTs \cite{Banados:1998gg,Roberts:2012aq,Shimaji:2018czt}. Below we extend it to that for a dS$_3$ and analyze the holographic pseudo entropy by using this map.

\subsection{Ba\~{n}ados-like map on dS$_3$}
We introduce the Ba\~{n}ados-like map to transform the perturbed dS$_3$ whose coordinates are $(\tau,w,\bar{w})$ into the Poincar\'{e} dS$_3$ whose coordinates are $(T,\xi,\bar{\xi})$, where $w=x+iy, \; \bar{w}=x-iy, \; \xi=X+iY, \; \bar{\xi}=X-iY$. 
This map can be obtained by the analytic continuation from Ba\~{n}ados map on Euclidean AdS$_3$ \cite{Banados:1998gg,Roberts:2012aq,Shimaji:2018czt}. \\
When we consider the following conformal transformation:
\be
\xi=f(w), \; \bar{\xi}=\bar{f}(\bar{w}),
\label{eq:conftrsf}
\ee
the corresponding Ba\~{n}ados map is given by 
\ba
&& \xi=f(w)+\frac{2\tau^2(f')^2(\bar{f}'')}{4|f'|^2-\tau^2|f''|^2},\no
&& \bar{\xi}=\bar{f}(\bar{w})+\frac{2\tau^2(\bar{f}')^2(f'')}{4|f'|^2-\tau^2|f''|^2},\no
&& T=\frac{4\tau(f'\bar{f}')^{3/2}}{4|f'|^2-\tau^2|f''|^2}.
\label{dSbanadosmap}
\ea
In this map, the Poincar\'{e} dS$_3$ metric $ds^2= R_{dS}^2 \frac{-dT^2+d\xi d\bar{\xi}}{T^2}$ changes as follows:
\ba
ds^2=-\frac{d\tau^2}{\tau^2}-T(w)(dw)^2-\bar{T}(\bar{w})(d\bar{w})^2+\left(\frac{1}{\tau^2}
+\tau^2T(w)\bar{T}(\bar{w})\right)dwd\bar{w},
\label{banadosmetric}
\ea
where
\ba
T(w)=\frac{3(f'')^2-2f'f'''}{4f'^2},\ \ \bar{T}(\bar{w})=\frac{3(\bar{f}'')^2-2\bar{f}'\bar{f}'''}{4\bar{f}'^2}.
\ea

\subsection{$\Delta S_A$ computed via Ba\~{n}ados-like map}

Now we would like to compute the entanglement entropy in the perturbed setup. We choose the subregion $A$ whose end points are $(w_1,\bar{w_1})$ and $(w_2,\bar{w_2})$, which is the same one as in section \ref{sec:cftt}.
Considering the infinitesimal conformal transformation
\ba
    \xi=w+\delta f(w), \qquad \bar{\xi}=\bar{w}+\delta \bar{f}(\bar{w}),
\ea
we can calculate the following entanglement entropy through the Poincar\'{e} dS$_3$ coordinate:
\ba
    S_A=&&\frac{i\tilde{c}}{6}\log\frac{(X_2-X_1)^2}{\ep_1\ep_2}+\frac{\pi \tilde{c}}{6} \\
    \simeq&& \frac{i\tilde{c}}{6}\log\frac{(x_2-x_1)^2}{\ep_{p1}\ep_{p2}}+\frac{\pi \tilde{c}}{6} +\frac{i\tilde{c}}{6}\Bigg(\frac{(\delta f(w_2)+\delta \bar{f}(\bar{w}_2))-(\delta f(w_1)+\delta \bar{f}(\bar{w}_1))}{x_2-x_1} \\
    &&-\frac{1}{2}\left(\delta f'(w_2)+\delta \bar{f}'(\bar{w}_2)+\delta f'(w_1)+\delta \bar{f}'(\bar{w}_1)\right)\Bigg),
\ea
where the UV cutoff $\ep_p$ in perturbed coordinate corresponds to the UV cut off in the Poincar\'{e} coordinate. \\
We immediately read off the perturbation part
\ba
    \Delta S_A \simeq \frac{i\tilde{c}}{6}\Bigg(\frac{\delta f(w_2)+\delta \bar{f}(\bar{w}_2)-\delta f(w_1)-\delta \bar{f}(\bar{w}_1)}{x_2-x_1}
     -\frac{\delta f'(w_2)+\delta \bar{f}'(\bar{w}_2)+\delta f'(w_1)+\delta \bar{f}'(\bar{w}_1)}{2}\Bigg), \notag \\
    \label{eq:perturbeddSDeltaS}
\ea
which coincides with \eqref{eq:dSDeltaS}. \\
% We use the holographic stress tensor $T_{\mu\nu}=\frac{iR_{dS}}{8\pi G_N}H_{\mu\nu}$, where $H_{\mu\nu}$ is the first order perturbation of the metric, obtained via analytic continuation from the AdS/CFT.
%Next, we can rewrite \eqref{banadosmetric} at the first order perturbation. 

Assuming $y$ is the Euclidean time coordinate in the dual CFT, we obtain the holographic energy stress tensor
\ba
    T_{yy} \simeq -\frac{i R_{dS}}{16\pi G_N}(\delta f'''(w)+\delta \bar{f}'''(\bar{w}))\simeq-\frac{i \tilde{c}}{24}(\delta f'''(w)+\delta \bar{f}'''(\bar{w})),
\ea
which agrees with \eqref{perturbedenergystresstensordS}. By combining these with the calculation in (\ref{DeltaEA}), this derived the first law in dS$_3$/CFT$_2$ correspondence using the Ba\~{n}ados map.\\

\section{Comparison of the energy stress tensor in dS$_3$/CFT$_2$}\label{ap:EM}
From \eqref{metricsolution} and \eqref{HJm}, we obtain the holographic energy stress tensor
\ba
    T_{\theta\theta}=&&\frac{iR_{dS}}{16\pi G_N}\frac{e^{im\phi}}{(\sin\theta)^2}\left[A_1\left(\tan\frac{\theta}{2}\right)^m
+A_2\frac{1}{\left(\tan\frac{\theta}{2}\right)^{m}}\right]  \label{Tthetatheta}\\
    T_{\theta\phi}=&&\frac{iR_{dS}}{16\pi G_N}\frac{ie^{im\phi}}{\sin\theta}\left[A_1\left(\tan\frac{\theta}{2}\right)^m
-A_2\frac{1}{\left(\tan\frac{\theta}{2}\right)^{m}}\right] \label{Tthetaphi} \\
    T_{\phi\phi}=&&\frac{-iR_{dS}}{16\pi G_N}e^{im\phi}\left[A_1\left(\tan\frac{\theta}{2}\right)^m
+A_2\frac{1}{\left(\tan\frac{\theta}{2}\right)^{m}}\right]. \label{Tphiphi}
\ea
We find that they have the singularities at $\theta=0,\,\pi$. These singularities normally appear in the spacetime with the compact boundary, such as Euclidean dS$_3$ and Euclidean AdS$_3$. They can be understood from the CFT$_2$ perspective as we will see below.  \\
Let us consider the operators which has the conformal weight $(h,\bar{h})$. Now we would like to calculate the energy stress tensor on the CFT$_2$ inserting two operators. Using the Laurent expansion of stress tensor, we obtain the expectation values
\ba
    \braket{O|T(z)|O}=\braket{O|z^{-2}L_0|O}=\frac{h}{z^2}
     \label{TOOprimary},
\ea
\ba
    \braket{O|T(z) (L_{-1})^m|O}=&&\frac{(m+1)!h}{z^{2+m}},\no
    \braket{O|(L_{1})^m T(z)|O}=&&(m+1)!hz^{-2+m},
    \label{TOOdescendants}
\ea
where $O$ is the primary operator and $(L_{-1})^mO$ are descendants.
We use the following stereographic projection to transform the planar coordinate to the spherical coordinate:
\ba
    z=\frac{e^{i\phi}\sin\theta}{1-\cos\theta}, \; \bar{z}=\frac{e^{-i\phi}\sin\theta}{1-\cos\theta}.
\ea
Then, we obtain
\ba
    \braket{O|T_{\theta\theta}|O} = \Bigg[\left(\frac{e^{i\phi}}{1-\cos\theta}\right)^2\braket{O|T(z)|O}+\left(\frac{e^{-i\phi}}{1-\cos\theta}\right)^2\braket{O|\bar{T}(\bar{z})|O}\Bigg] = \frac{h+\bar{h}}{\sin^2\theta},\no
    \label{EVTthetatheta_primary}
\ea
and similarly
\ba
    \braket{O|T_{\theta\theta}(L_{-1})^m|O} =&& \frac{(m+1)!(h+\bar{h})}{\sin^2\theta}(e^{im\phi}+e^{-im\phi})\left(\tan\frac{\theta}{2}\right)^m, \notag \\
    \braket{O|(L_{1})^m T_{\theta\theta}|O} =&& \frac{(m+1)!(h+\bar{h})}{\sin^2\theta}(e^{im\phi}+e^{-im\phi})\frac{1}{\left(\tan\frac{\theta}{2}\right)^m}.
    \label{EVTthetatheta_descendants}
\ea
Focusing on the $\theta$ and $\phi$ dependence, the first equation and the second equations respectively correspond to  $m$ and $-m$ modes of \eqref{Tthetatheta}. We can similarly calculate the expectation values of $T_{\theta\phi}$ and $T_{\phi\phi}$ and find they have the same dependence as that of \eqref{Tthetaphi} and \eqref{Tphiphi}.

\section{First law in Poincar\'e coordinate}\label{ap:poin}

\begin{figure}
    \centering
    \includegraphics[width=0.7\linewidth]{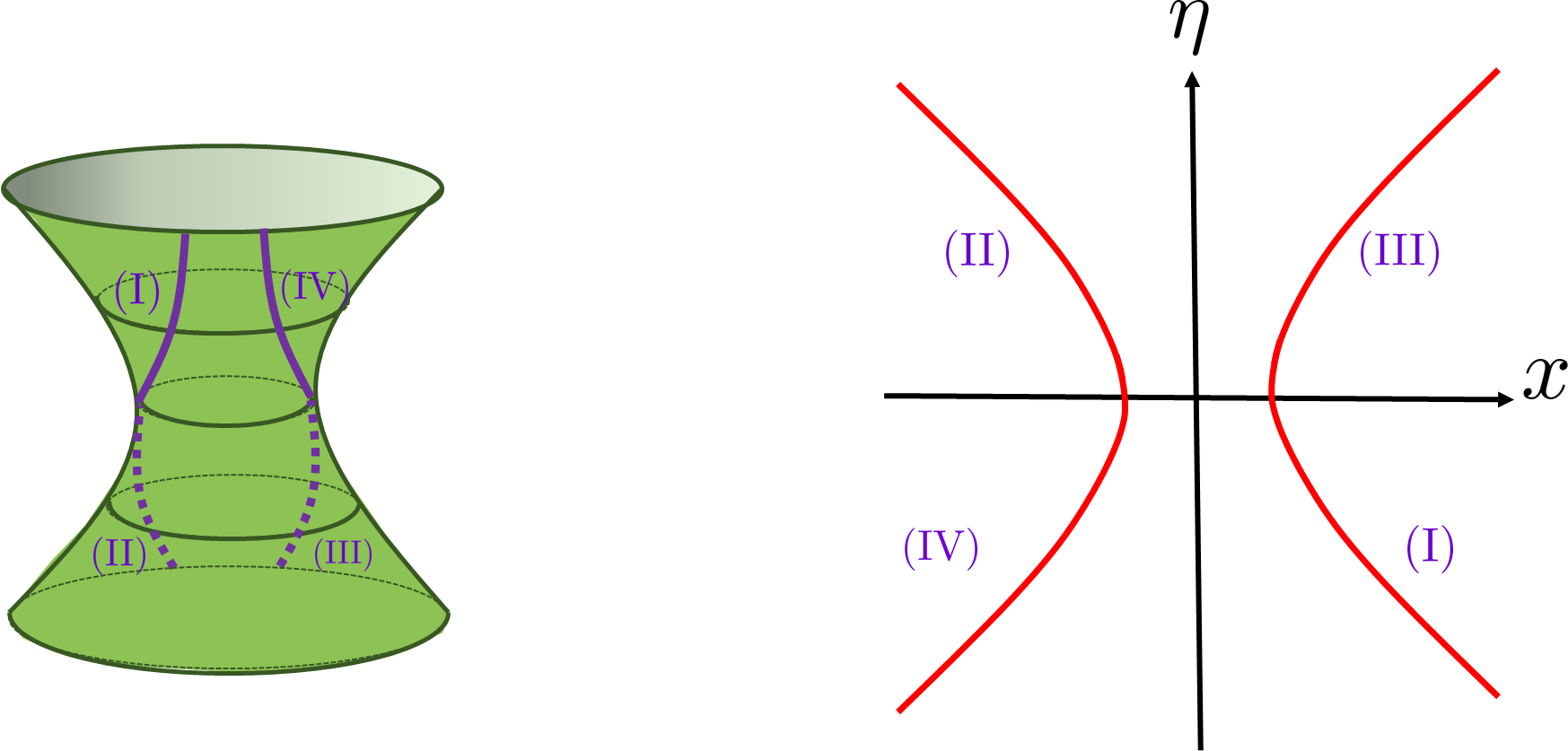}
    \caption{dS geodesics in global patch and Poincar\'e patch.  To get the whole length of half Lorentzian region, we have to calculate (I)$+$(IV).}
    \label{fig:dS_geodesics_global_Poincare}
\end{figure}

In this section, we consider how the analytic continuation from the result in AdS$_3$ Poincar\'e patch \cite{Nozaki:2013vta} is realized in dS$_3$ Poincar\'e patch.
Also, we comment the relation to holographic timelike entanglement investigated in \cite{Heller:2024whi}. 

As shown in Fig.\ref{fig:dS_geodesics_global_Poincare}, the lower half of the Poincar\'e patch ($\eta<0$) covers the future half region in dS$_3$ and contains geodesic segments (I) and (IV), whereas the upper half ($\eta>0$) covers the past region, comprising segments (II) and (III).
In this patch, we introduce the perturbation
\begin{align}
    g_{\alpha\beta}\mathrm{d}x^\alpha\mathrm{d}x^\beta=R_{\text{dS}}^2\frac{-\mathrm{d}\eta^2+(\delta_{\mu\nu}+h_{\mu\nu})\mathrm{d}x^\mu\mathrm{d}x^\nu}{\eta^2},
\end{align}
and calculate $\Delta S_A$.

\begin{figure}
    \centering
    \includegraphics[width=0.8\linewidth]{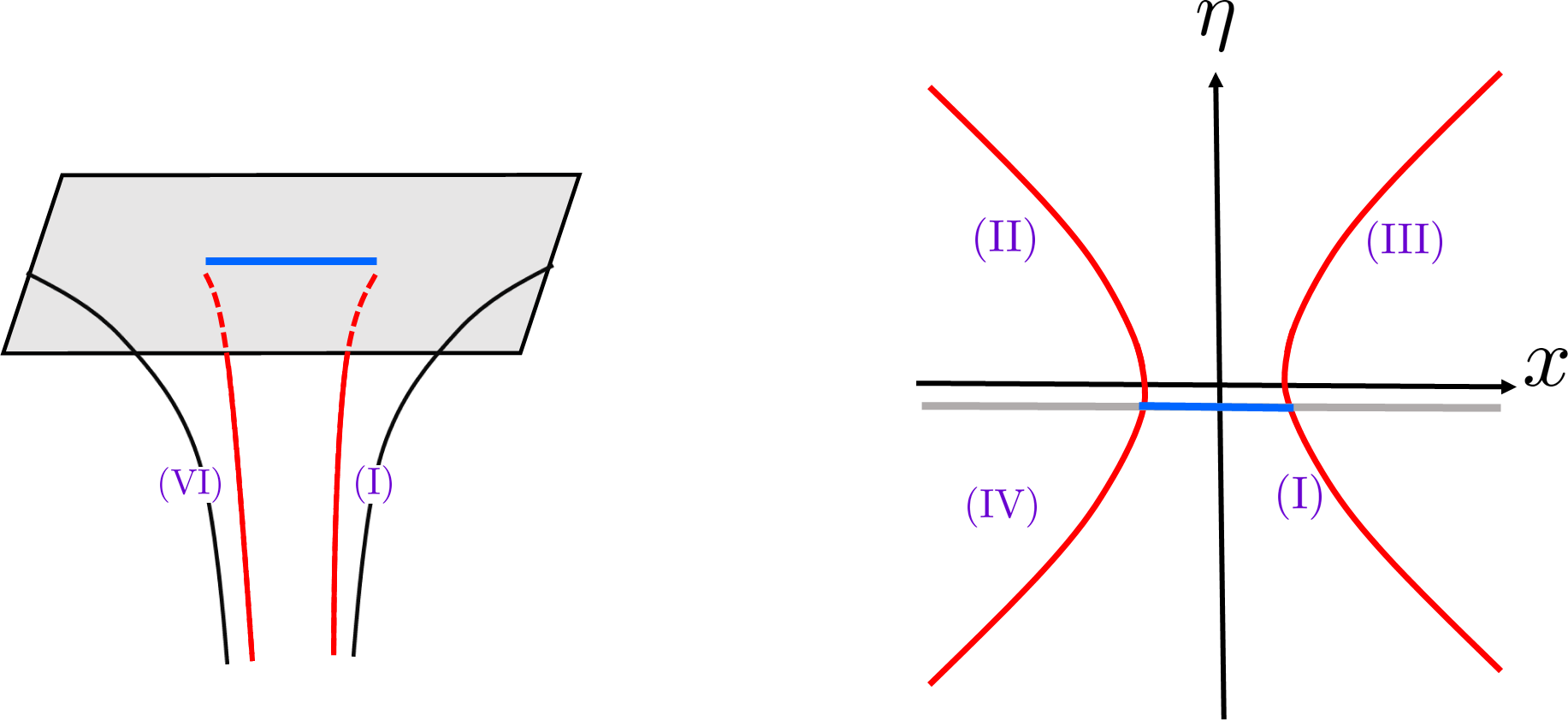}
    \caption{dS$_3$ geodesic in Poincar\'e patch.}
    \label{fig:geodesic_Poincare}
\end{figure}
Consider the boundary interval at $y=0$ with $-\frac{l}{2}<x<\frac{l}{2}$ and impose the UV cutoff at $\eta=\epsilon$. 
The geodesic equation then reads
\begin{align}
    x^2-\eta^2=\frac{l^2}{4},
\end{align}
where $(x,\eta)$ can be parametrized by $u$ using hyperbolic functions:
\begin{align}
    x=\frac{l}{2}\cosh{u},\quad y=-\frac{l}{2}\sinh{u}.
\end{align}
Thus we can calculate geodesic as $u$ integral:
\begin{align}
    L_{\text{Lorentz}}&=iR_{\text{dS}}\int_{+u_\epsilon}^{+\infty}\frac{\mathrm{d}u}{\sinh u}+iR_{\text{dS}}\int^{-u_\epsilon}_{-\infty}\frac{\mathrm{d}u}{\sinh (u-i\pi)}\notag\\
    &=2iR_{\text{dS}}\log{\left(\frac{l}{\epsilon}\right)}.
\end{align}
where $\sinh u_{\epsilon}=\frac{2\epsilon}{l}$.

\begin{figure}
    \centering
    \includegraphics[width=0.7\linewidth]{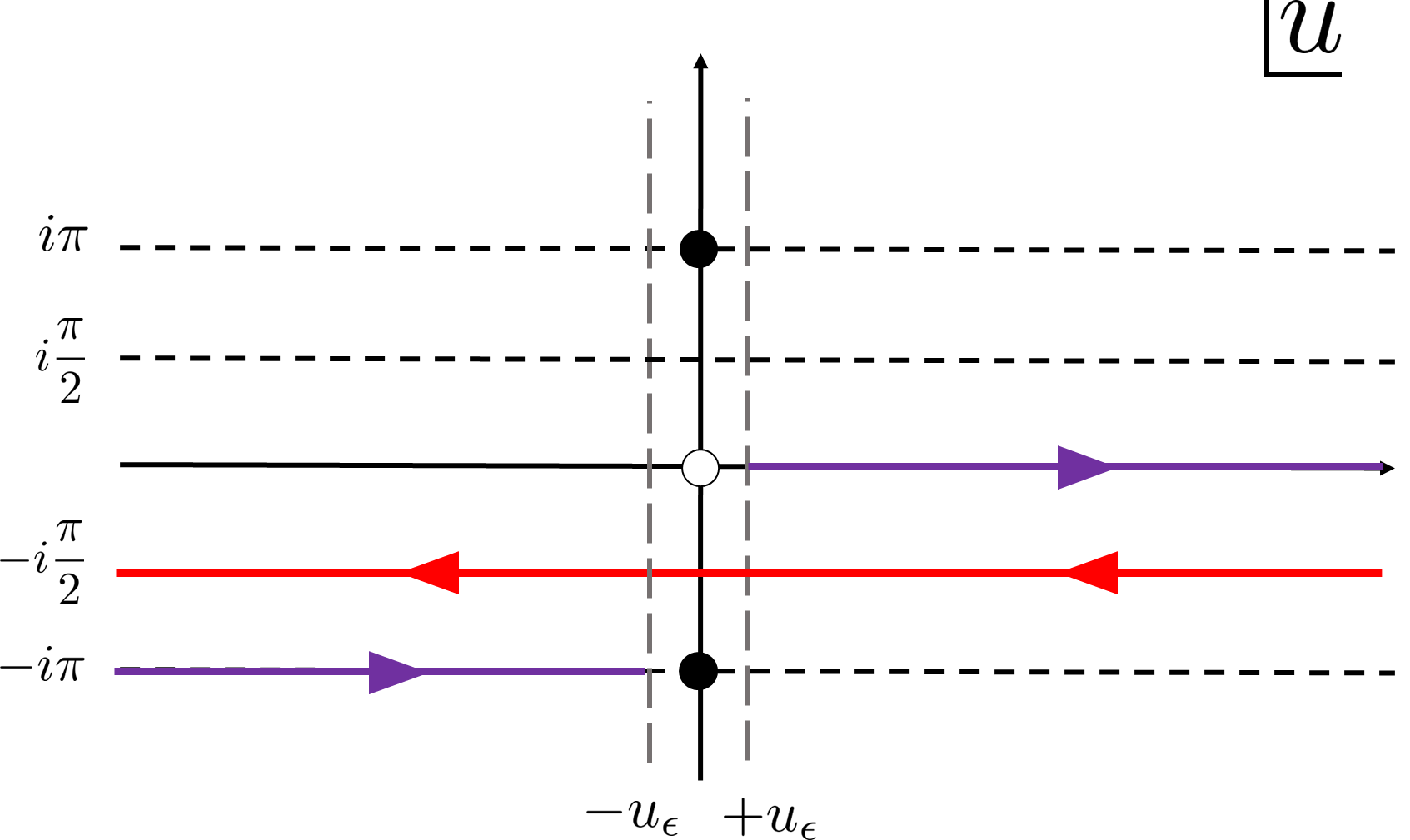}
    \caption{Complex contour of integration $\frac{1}{\sinh u}$. White poles are corresponding to the poles whose residue are $+2i \pi$ while black ones are $-2i\pi$.}
    \label{fig:complex _contour_of_u}
\end{figure}

\begin{figure}
    \centering
    \includegraphics[width=0.5\linewidth]{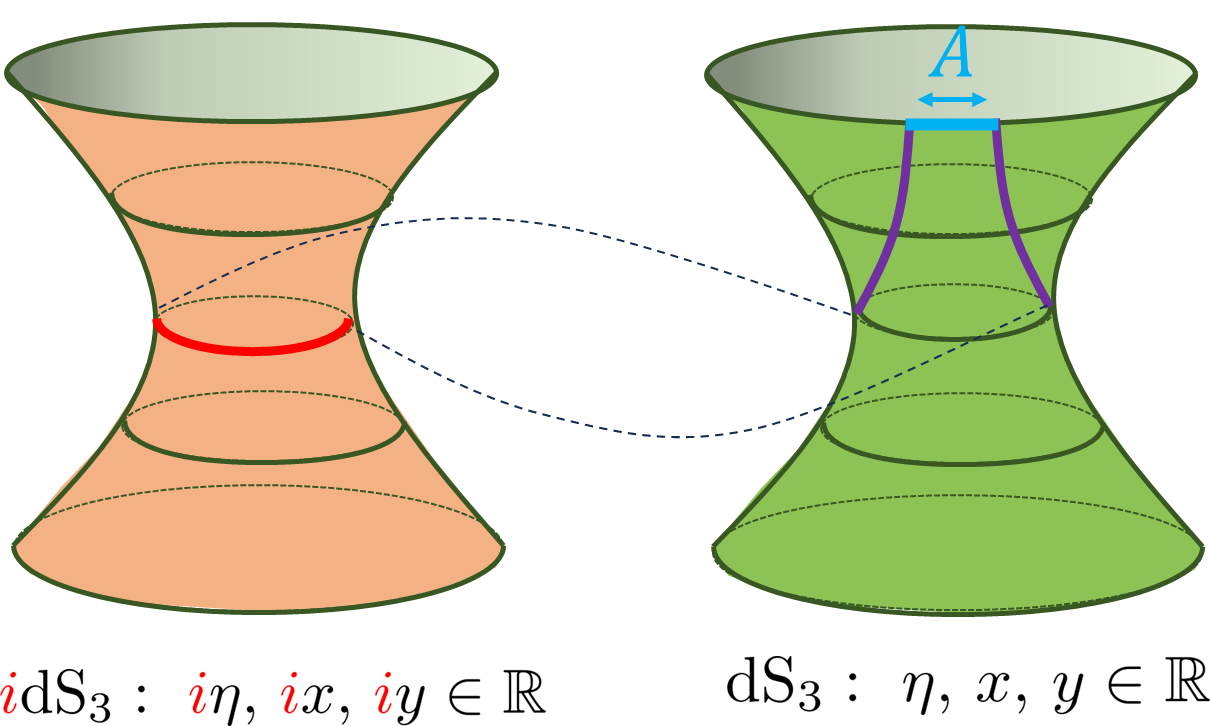}
    \caption{The contour in Fig \ref{fig:complex _contour_of_u} can be interpreted as a geodesics in complexified dS$_3$.}
    \label{fig:complex_Pincare}
\end{figure}

As we did in section \ref{sec:dsCFTF}, we have to connect the geodesic using complex contour of $u$.
Then, we find the red line in Fig \ref{fig:complex _contour_of_u} is one of the contour connecting $u=+\infty$ and $u=i\pi-\infty$, where the whole length is obtained:
\begin{align}
    L=iR_{\text{dS}}\int_{\text{purple+red}}\frac{\mathrm{d}u}{\sinh u}=2iR_{\text{dS}}\log{\left(\frac{l}{\epsilon}\right)}+\pi.
\end{align}
The contour in Fig \ref{fig:complex _contour_of_u} can be interpreted as a geodesics in complexified dS$_3$ represented by Fig \ref{fig:complex_Pincare}.
This is very similar to the geodesics in complexified AdS$_3$ dual to the timelike entanglement entropy \cite{Heller:2024whi}.

\begin{figure}
    \centering
    \includegraphics[width=0.7\linewidth]{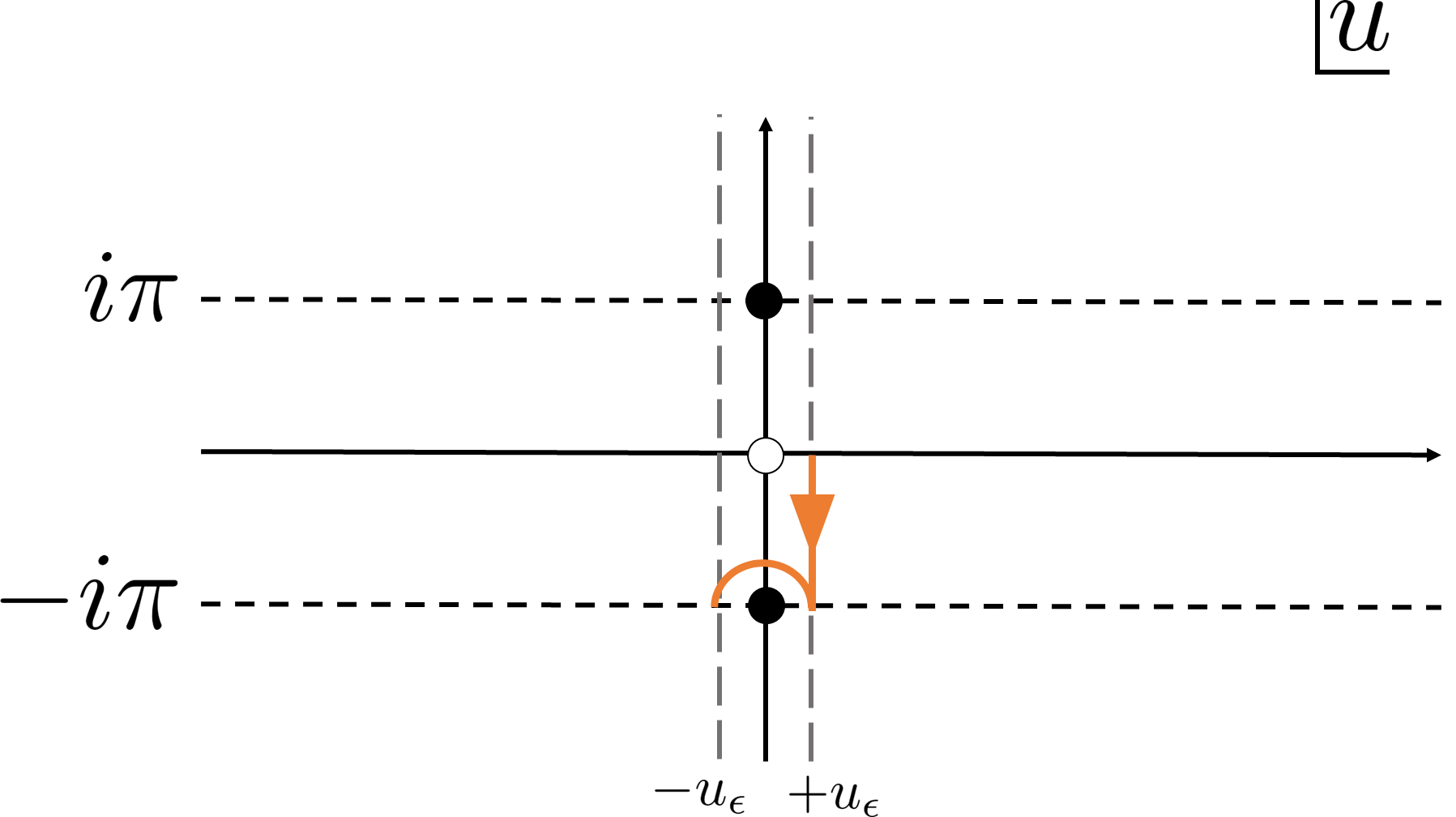}
    \caption{Complex contour of integration $\frac{1}{\sinh u}$ equivalent to Fig \ref{fig:complex _contour_of_u}. White poles are corresponding to the poles whose residue are $+2i \pi$ while black ones are $-2i\pi$.}
    \label{fig:complex _contour_of_u2}
\end{figure}
Alternatively, we can deform the contour to the orange path shown in Fig \ref{fig:complex _contour_of_u2}, which yields the same result as Fig \ref{fig:complex _contour_of_u}.
If we use the orange contour in Fig \ref{fig:complex _contour_of_u2} calculating $\Delta S_A$, we find the result is nothing but the analytic continuation from the Poincar\'e AdS$_3$ \cite{Nozaki:2013vta}:
\begin{align}
    \Delta S_A=&\frac{i R_{\text{dS}} l^2}{32 G_N}\int_{0}^{\pi}(i\mathrm{d}v)(-i\cosh u_{\epsilon}\sin v)^3H\left(\xi+\frac{l}{2}\cosh{u_\epsilon}\cos v,y\right)\notag\\
    =&\frac{i R_{\text{dS}} l^2}{32 G_N}\int_{0}^{\pi}(i\mathrm{d}v)(-i\sin v)^3H\left(\xi+\frac{l}{2}\cos v,y\right)\notag\\
    =&-\frac{i R_{\text{dS}} l^2}{32 G_N}\int_{0}^{\pi}\mathrm{d}v\sin^3 v\  H\left(\xi+\frac{l}{2}\cos v,y\right),
\end{align}
where 
\begin{align}\label{perturbation_hxy}
    v=\text{Im}{(u)},\quad
    h_{xx}=h_{yy}=\eta^2\,H(x,y),
\quad
\partial_{x}h_{xy}=-\eta^2\,\partial_{y}H,
\quad
\partial_{y}h_{xy}=\eta^2\,\partial_{x}H.
\end{align}
We note the function $H$ satisfies 
\begin{align}
    (\partial_{x}^{2}+\partial_{y}^{2})\,H(x,y)=0.
\end{align}

By taking the Fourier transformation
\begin{align}
    H(k,y)=\int_{-\infty}^{+\infty}\mathrm{d}x\ e^{-ikx}\,H(x,y),
\end{align}
we obtain
\begin{align}
    \Delta S_A(k,l,y)=\frac{-iR_{\text{dS}}}{2G_N}\frac{2\sin \frac{kl}{2}-lk\cos \frac{kl}{2}}{k^3l}H(k,y)
\end{align}
and we find the EOM in the momentum space of $\xi$:
\begin{align}
    \left[\partial_l^2+\left(\frac{k^2}{4}-\frac{2}{l^2}\right)\right]\Delta S_A=0.
\end{align}
This leads to the EOM in coordinate space:
\begin{align}
    &(\partial_\xi^2+\partial_y^2)\Delta S_A(\xi,l,y)=0,\\
    & \left[\partial_l^2+\frac{1}{4}\partial_y^2-\frac{2}{l^2}\right]\Delta S_A(\xi,l,y)=0.
\end{align}
These equations are easily obtained via the analytic continuation from the result in Poincar\'e AdS$_3$ \cite{Nozaki:2013vta}.

%%%%%%%%%%%%%%%%%%%%%%%%%%%%%%%%%%%%%%%%%%%%%%%%%%%%%%%%
%%%%%%%%%%%%%%%%%%%%%%%%%%%%%%%%%%%%%%%%%%%%%%%%%%%%%%%%
% bibliography via BibTeX
\bibliographystyle{JHEP}
\bibliography{dS_Einstein}

%%%%%%%%%%%%%%%%%%%%%%%%%%%%%%%%%%%%%%%%%%%%%%%%%%%%%%%%
%%%%%%%%%%%%%%%%%%%%%%%%%%%%%%%%%%%%%%%%%%%%%%%%%%%%%%%%

\end{document}